\documentclass[journal=jctcce,manuscript=article,maxauthors=0,hyperref=true]{achemso}
\setkeys{acs}{articletitle=true}

\pdfoutput=1

\usepackage{color}
\usepackage{braket}
\usepackage{mathtools}
\usepackage[version=3]{mhchem} 
\usepackage{threeparttable}
\usepackage{tablefootnote}
\usepackage{siunitx}
\usepackage{gensymb}
\usepackage{textcomp}

\DeclarePairedDelimiter\abs{\lvert}{\rvert}

\definecolor{acolour}{RGB}{153, 76, 0}

\newcommand{\mr}{\mathrm}

\newcommand{\mi}{\mu^i}
\newcommand{\qi}{\theta^i}
\newcommand{\oi}{\Omega^i}
\newcommand{\hi}{\Phi^i}
\newcommand{\dmi}{\Delta\mu^i}
\newcommand{\dqi}{\Delta\theta^i}

\newcommand{\oil}{\Omega^{i'}}
\newcommand{\hil}{\Phi^{i'}}

\newcommand{\ddl}{\alpha^{i'}}
\newcommand{\dql}{A^{i'}}
\newcommand{\qql}{C^{i'}}

\newcommand{\al}{\alpha}
\newcommand{\ab}{{\alpha\beta}}
\newcommand{\aby}{{\alpha\beta\gamma}}
\newcommand{\abyd}{{\alpha\beta\gamma\delta}}
\newcommand{\abyde}{{\alpha\beta\gamma\delta\epsilon}}

\newcommand{\be}{\beta}
\newcommand{\by}{{\beta\gamma}}
\newcommand{\byd}{{\beta\gamma\delta}}
\newcommand{\byde}{{\beta\gamma\delta\epsilon}}

\newcommand{\br}{{\bf r}}

\newcommand{\ecom}{COM}

\author{Elvar \"Orn J\'onsson}
\email{eojons@gmail.com}
\affiliation[University of Iceland]{Science Institute and Faculty of Physical Sciences, University of Iceland, VR-III, 107 Reykjavík, Iceland}
\author{Soroush Rasti}
\affiliation[Leiden University]{Leiden Institute of Chemistry, Gorlaeus Laboratories, Leiden University, 2300 RA Leiden, The Netherlands}
\author{Marta Galynska}
\affiliation[University of Iceland]{Science Institute and Faculty of Physical Sciences, University of Iceland, VR-III, 107 Reykjavík, Iceland}
\author{J\"org Meyer}
\affiliation[Leiden University]{Leiden Institute of Chemistry, Gorlaeus Laboratories, Leiden University, 2300 RA Leiden, The Netherlands}
\author{Hannes J\'onsson}
\affiliation[University of Iceland]{Science Institute and Faculty of Physical Sciences, University of Iceland, VR-III, 107 Reykjavík, Iceland}

\title{Transferable Potential Function \\ for Flexible H$_2$O 
Molecules \\ Based on the Single Center Multipole Expansion}

\abbreviations{MM}

\begin{document}

\begin{tocentry}
\includegraphics[width=8.255cm]{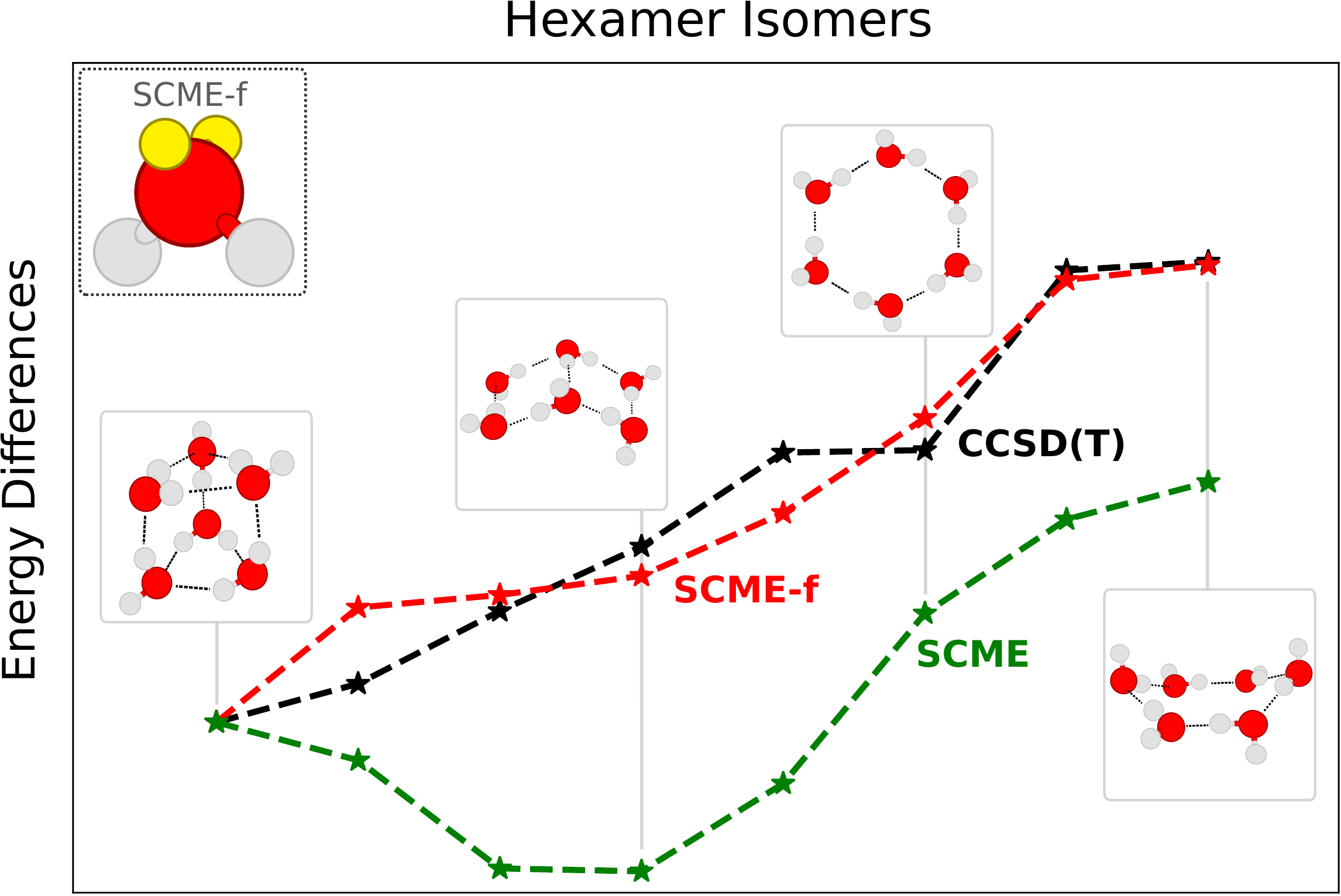}
\end{tocentry}

\begin{abstract}
A potential function is presented for
describing a system of flexible H$_2$O molecules based on the single center multipole
expansion (SCME) of the electrostatic interaction. 
The model, referred to as SCME/f,
includes the variation of the molecular quadrupole moment as well as the dipole moment
with changes in bond length and angle  
so as to reproduce results of high level 
electronic structure calculations.
The multipole expansion also includes
fixed octupole and hexadecapole moments,  
as well as
anisotropic dipole-dipole, dipole-quadrupole
and quadrupole-quadrupole polarizability tensors.
The model contains 
five adjustable parameters related to the repulsive interaction and damping functions in the electrostatic and dispersion interactions.
Their values are adjusted to reproduce 
the lowest energy isomers 
of small clusters, (H$_2$O)$_n$ with $n=2-6$, as well as 
measured properties of the ice Ih crystal. 
Subsequent calculations of the energy difference between the various
isomer configurations of the clusters show that
SCME/f gives 
good agreement with results of electronic structure
calculations and represents a significant improvement over the previously presented rigid SCME potential function. 
Analysis of the vibrational frequencies of the clusters
and structural properties of ice Ih crystal
show the importance of accurately describing 
the variation of the quadrupole moment with molecular structure.

\end{abstract}



\section{\label{sec: introduction} Introduction}

The most commonly used potential energy functions for describing
water molecules and their interaction are based on simple pairwise additive functions
with fixed point charges \cite{TIP3P,TIP4P,horn:2004, Zielkiewicz2005},
such as the well known TIPnP and SPC 
force fields. 
Extensions of these potential functions to 
describe 
flexible molecules 
have been developed,
such as aSPC/Fw\cite{aspcfw} and q-TIP4P/F\cite{qtip4pf}, and they offer, for example, 
the possibility to include 
the effect of zero point energy.
The point charge potential functions are typically parameterized in such a way as to reproduce 
a few thermally averaged properties 
of liquid water.
The properties of water molecules are, however, strongly environment dependent as illustrated by the 
molecular dipole moment, which is 1.8 D 
in the gas phase and 3.1 D in ice Ih \cite{batista:1998}. This large environment dependence
needs to be modeled accurately in order to
develop a transferable potential function applicable, for example, to small clusters and crystal structures
as well as liquid water.

Such environment dependence is best described using well established physical laws, since empirical fitting to some limited set of data is likely not going to work well
when the potential function is applied to configurations that are significantly different from the ones used in the fitting process.
A systematic multipole expansion up to and including the hexadecapole,
with dipole and quadrupole polarizability, 
has been shown to reproduce well the electrostatics in water clusters and ice
\cite{batista:2000}.
A potential function based on this approach has been presented for rigid molecules
and is referred to as
the single center multipole expansion (SCME) potential function \cite{scme,wikfeldt:2013}.
In the present work, this approach is extended to flexible molecules.

By expanding the electrostatics around a single center on each molecule, 
the introduction of point charges is avoided
and the correct long range distance 
dependence of the Coulomb potential built in naturally.
The leading term, the dipole potential, decays as $1/R^3$, 
and combined with the polarization response of the molecules 
this makes it possible to
use a long range cut-off
for the electrostatic interaction between molecules in typical condensed matter simulations.
\cite{batista:1998}.

Hybrid simulation schemes, where part of the system is simulated using a potential function while another part is described 
using electronic structure calculations, the so-called
quantum mechanics / molecular mechanics (QM/MM) simulations, 
have been used in important simulation studies in various fields such has 
biochemistry\cite{smirnov:2016, barends:2016, senn:2009, senthilkumar2008analysis, warshel:2006}, 
medicine \cite{zheng:2014}, photochemistry\cite{knorr:2016} and solvation dynamics\cite{pham:2011, dohn:2014, dohn:2016, levi:2018},
nanostructures\cite{dohn:2018}, and materials science.\cite{zhang2018}
In most cases, such simulations
make use of 
fixed point charge models 
\cite{lin:2006, pezeshki:2015, sneskov2011scrutinizing, Morzan2018},
thereby neglecting the 
mutual
polarization of the 
charges in the MM subsystem by the QM subsystem -- an effect that was,
however,
included in the inceptive work initiating the QM/MM approach\cite{Warshel1976}.
The use of fixed point charge models to represent water molecules in the MM region 
results in errors that limit the applicability of the QM/MM method. 

Several H$_2$O potential functions that include some level of polarizability 
exist\cite{cisneros:2016,yu2005accounting,lopes2009molecular}. 
These 
include the Thole-type multipole models such as the TTMn 
series\cite{burnham2002development,ttmn2fr,fanourgakis2008development,burnham2008vibrational}, 
and HBB2-pol\cite{hbbpola,hbbpolb}. 
The MB-pol\cite{mbbpola,mbbpolb,mbbpolc} potential function has arguably 
reached the highest precision as it includes an explicit 
treatment of two-body and three-body interactions through an intricate permutationally 
invariant polynomial fit to data bases constructed with high level quantum chemistry calculations.
However, inclusion of such explicit many body terms makes the interfacing
with a QM region more challenging. Instead, simpler polarizable MM potential functions
based on pair-wise potentials to describe the short-range interactions are used in so-called polarizable embedding QM/MM 
(PE-QM/MM)
approach
\cite{thompson1995excited, doi:10.1021/jp960690m, BRYCE1997367, lipparini2011polarizable, doi:10.1021/ct300722e, lu2008interfacing, thellamurege2013quanpol, kratz2016lichem, dziedzic2016tinktep, gomes2012quantum, soderhjelm2009protein, sneskov2011scrutinizing, sneskov2011polarizable, caprasecca2014geometry, kongsted2002qm, zeng2015analytic, loco:2016, loco2017hybrid, jensen2003discrete, doi:10.1021/jp1101913, nielsen2007density, olsen2010excited, lipparini2012linear, doi:10.1021/ct9001366, list2016excited, schworer2013coupling,Curutchet2009,Visscher2018,Hrsak2018,Menger2017,mao2017,doi:10.1063/1.5080384}. 
The
PE-QM/MM 
approach
can be used to 
study the effects of solvation and solvent response to excitations 
and charge transfer in solvated species.
However, 
such simulations have typically included only the
molecular dipole-dipole response
and make use of atomic point charges.

Here, we describe an extension of the single-center multipole expansion \cite{scme,wikfeldt:2013} (SCME) 
potential function, which has recently been 
integrated in a 
PE-QM/MM
scheme \cite{peqmmm1,peqmmm2}. The extended potential function, SCME/f, includes flexibility of the
internal geometry of the water molecules while still maintaining 
the single center description of the electrostatic 
interaction in terms of molecular moment tensors.
The SCME/f model includes variable 
dipole and quadrupole moment
tensors 
that depend on the geometry of the H$_2$O molecule.
The dipole is described by the well established Partridge-Schwenke 
model,\cite{partridge1997}
but a new, geometrical model based on four sites is presented here 
for the quadrupole moment.
It reproduces results of high-level multireference electronic structure 
calculations of the quadrupole moment to within 1.6\% RMS over a broad range in its magnitude.
This model for the quadrupole moment 
is found to provide better description than the  
so-called M-site models
that 
have been used previously.\cite{TIP4P,scott2009,horn2004development,abascal2005potential,abascal2005general,burnham2002development,ttmn2fr,fanourgakis2008development,burnham2008vibrational,hbbpola,hbbpolb,mbbpola,mbbpolb,mbbpolc}

There are five 
adjustable
parameters 
in the description of the 
intermolecular interaction. They include 
parameters relating to the pair-wise repulsive interaction as well as damping parameter in the
dispersion interaction and
a screening parameter for the 
electrostatic interaction tensors.
These parameters are optimized in such a way that the SCME/f
reproduces the binding energy and intermolecular distance of the 
dimer, the
interaction energy of the lowest energy conformation of water clusters (H$_2$O)$_n$ with $n$ ranging from 3 to 6, 
calculated at the level of RI-MP2 with CCSD(T) corrections\cite{bates:2009} and full CCSD(T) at the complete basis set limit\cite{temelso:2011}
as well as 
measured 
properties of crystalline 
ice Ih 
taking into account the
zero-point energy.
The resulting
parametrization of the model reproduces nicely trends 
in the relative energy of the 
conformers of the hexamer
obtained from high level 
quantum chemistry calculations.
Some
discrepancies, however, exist in the series of pentamer isomers. 
An analysis of the frequency of vibrational modes of the 
various
clusters 
and the structure of ice Ih crystal
highlights the importance of 
an accurate model for the molecular quadrupole moment.

The article is organized as follows: 
The SCME/f potential function is described in Section
\ref{sec:theory}.
The dipole and quadrupole surfaces are presented in Section
\ref{sec:dms_qms} and the calculation of
atomic forces is described in Section \ref{subsec: ana_for}.
The fitting of the five adjustable 
parameters is described in Section~\ref{sec:model_fit} and 
comparison with {\it ab initio} data on the cluster conformer energy and vibrational 
frequencies of small clusters is described in Section
\ref{sec:validation}.
Discussion and conclusions are in Section \ref{sec:discussion}.


\section{\label{sec:theory} Flexible SCME Model}

\begin{figure}[!th]
\includegraphics[width=.35\textwidth]{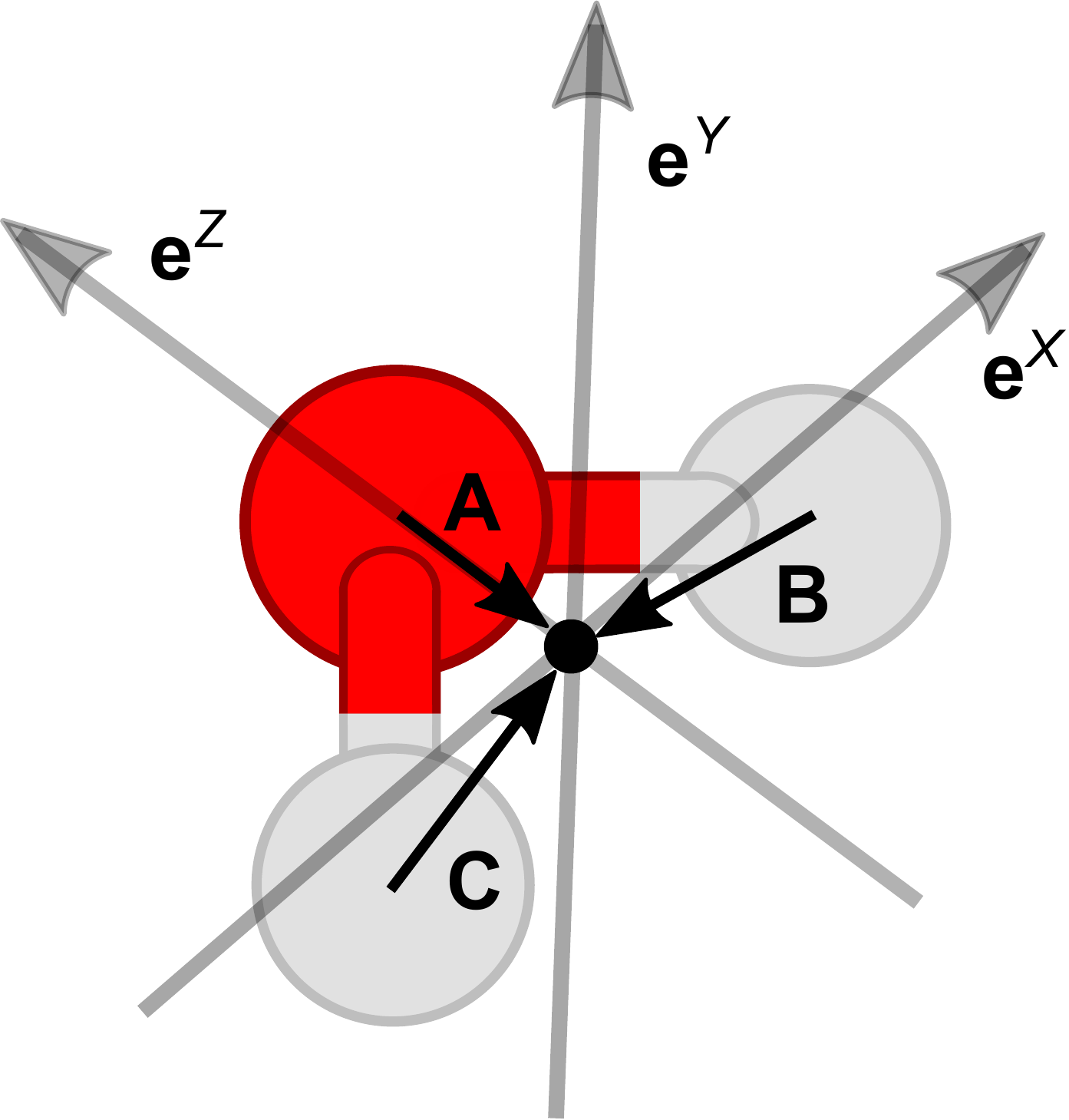}
\caption{%
The definition of the principal vectors and local reference frame for the water molecule used in the SCME/f model. 
The black circle denotes the expansion center,
chosen here to be at the center of mass. 
Black arrows
show the three principal vectors ${\bf A}$, ${\bf B}$ and ${\bf C}$ pointing {\it from} the oxygen and the hydrogen atoms {\it to} the expansion center.
The gray opaque arrows show the local reference frame basis vectors $\{{\bf e}^X,{\bf e}^Y,{\bf e}^Z\}$. 
The principal vectors ${\bf B}$ and ${\bf C}$ define a local-to-global reference frame rotation matrix.
Due to symmetry specific indexing
of the atoms is omitted, and positions and scales are exaggerated for clarity.
}
\label{fgr:localframe}
\end{figure}

Figure \ref{fgr:localframe} shows the principal vectors which define both
the position of the expansion center and the local-to-global reference 
frame rotation matrix for the flexible
water molecule. The local frame origin is placed at the center of
mass (\ecom). In SCME/f each water molecule is ascribed a 
molecular dipole and quadrupole moments in terms
of variable partial charges based on the internal geometry, 
$\mu^i_\al(\{\br^{ia}\})$ and $\theta^i_\ab(\{\br^{ia}\})$, respectively, 
where 
$\{\br^{ia}\} = \{\br^{i\mr{O}},\br^{i\mr{H_1}},\br^{i\mr{H_2}}\}$, and is the set of position vectors for atoms $a$ in molecule $i$ in the 
global reference frame.
The details
of the dipole moment and quadrupole moment surfaces are described 
in section \ref{sec:dms_qms}. The index $i$ is used to denote both the specific water molecule,
as well as the corresponding \ecom\ site. Furthermore, each water 
molecule is ascribed, in the local reference frame, a fixed octupole,
$\oil_\aby$, and hexadecapole, 
$\hil_\abyd$, moment tensors, as well as polarizability tensors 
including dipole-dipole, $\ddl_\ab$, dipole-quadrupole, $\dql_\aby$, 
and quadrupole-quadrupole, $\qql_\abyd$, induction terms.

Lipparini et. al.\cite{lipparini2014scal} describe commonly used 
local reference frames and associated rotation matrices. 
The derivation here follows closely 
their work, with some obvious sign changes. The expansion center
is placed at the \ecom\ 
\begin{equation}
    \br^i = \sum_{a}^{n_i}\br^{ia} \frac{M^a}{M^i} \label{eq:com}
\end{equation}
where $n_i$ denotes the atomic sites $\{$O,H$_1$,H$_2\}$ of molecule $i$, and
$M^a$ and $M^i$ is the mass of the atom and molecule, respectively.
The principal vectors used to define the rotation are
\begin{equation}
    {\bf B}^i = \br^i - \br^{i{\mr{H}_1}},\ \ \ \
    {\bf C}^i = \br^i - \br^{i{\mr{H}_2}}
\quad ,
\end{equation}
where in general, i.e. for a flexible \ce{H2O} molecule, $B^i \neq C^i$.
Unit basis vectors are in terms of the principal vectors given by
\begin{align}
    {\bf e}^{iZ} =& 
    \frac{B^i{\bf C}^i + C^i{\bf B}^i}
    {\abs{B^i{\bf C}^i + C^i{\bf B}^i}} \nonumber \\
    {\bf e}^{iX} =& 
    \frac{{\bf B}^i - ({\bf B}^i\cdot {\bf e}^{iZ}){\bf e}^{iZ}}
    {\abs{{\bf B}^i - ({\bf B}^i\cdot {\bf e}^{iZ}){\bf e}^{iZ}}} \nonumber \\
    {\bf e}^{iY} =&  {\bf e}^{iZ}\times{\bf e}^{iX} \label{eq:cross}
\end{align}
where ${\bf e}^{iZ}$ is, as defined above, the bisector
between
the two oxygen-hydrogen bonds. In terms of the unit basis vectors a unitary local-to-global reference frame rotation matrix is
\begin{equation}
    {\bf R}^i = \begin{bmatrix}
             e^{iX}_x & e^{iX}_y & e^{iX}_z \\
             e^{iY}_x & e^{iY}_y & e^{iY}_z \\
             e^{iZ}_x & e^{iZ}_y & e^{iZ}_z
            \end{bmatrix} \label{eq:rot_mat}
\end{equation}
Given the rotation matrix for each molecule the fixed moment and polarizability matrices are 
rotated into the
global reference frame for each \ecom\ site $i$\footnote{Throughout this work we make use of Einstein notation, i.e. Cartesian vector
spaces are indexed with Greek letters, $\al=\be=\dots=\nu \in \{x,y,z\}$, and 
repeated Greek indices are to be summed over.}

\begin{equation}
    M^i_{\al\dots\delta} = R^i_{\eta\al}\dots R^i_{\sigma\delta}M^{i'}_{\eta\dots\sigma}
\end{equation}
where $M^{i}_{\al\dots\delta}$ is a generalized tensor of order $t$, requiring $t$ rotation operations (e.g. $\al^i_\ab = R^i_{\eta\al}R^i_{\tau\be}\ddl_{\eta\tau}$).
With the definitions above atomic forces are derived 
(see the Supplementary Information)
from the contribution of the fixed moments and polarizabilities 
to the electrostatic interactions involving
the single expansion center on each molecule.

General formulation, and notation, of the perturbative expansion 
of the electrostatic
intermolecular interaction -- resulting in the multipole moment 
model -- can be found 
elsewhere\cite{stone2013theory}. Here we only present the main
expressions
which are used to arrive at a self-consistent solution
to polarized 
molecular moments at sites $i$ in response to the external 
field due to all other neighboring molecules $j(\neq i)$. 

Given the external field, $V^i_\al$ (negative of the electric field),
and the field gradient, $V^i_\ab$, at the \ecom\ of $i$, the molecules are 
polarized resulting in
induced dipole and quadrupole moments
\begin{align}
\dmi_\al = -\al^i_\ab V^i_\be 
                - \frac{1}{3}A^i_{\aby}V^i_\by \label{eq:indpol} \\
\dqi_\ab = -A^i_{\gamma\ab}V^i_\gamma 
                   - C^i_{\gamma\delta\ab}V^i_{\gamma\delta}
                   \label{eq:indquad}
\end{align}
where the external field is given by
\begin{equation}
V^{i}_\al  = \sum^n_{j\neq i}V^{ij}_\al
\end{equation}
and the contribution to the external field at site $i$ due to
site $j$ is given by
\begin{align}
V^{ij}_\al =& -T_\ab^{ij}(\mu^j_\be(\{\br^{jb}\}) + \Delta\mu^j_\be) 
             + \frac{1}{3}T_\aby^{ij}(\theta^j_\by(\{\br^{jb}\}) + \Delta\theta^j_\by) 
             \nonumber \\
             &\ -\frac{1}{15}T_\abyd^{ij}\Omega^j_\byd
             +\frac{1}{105}T_\abyde^{ij}\Phi^j_\byde \label{eq:scmepot}
\end{align}
The field gradient -- and higher order gradients -- are given by 
the subsequent use of the gradient operator, $\nabla_\be V^i_\al = V^i_\ab$, $\nabla_\gamma V^i_\ab = V^i_\aby$.

At the start the
external field and field gradient due to the intrinsic moments is 
evaluated at each site. 
This results in an induced
dipole
and quadrupole moment, which in turn results in a change in the
external field and field gradient. A self-consistent solution to the non-linear relation 
between eqs~\eqref{eq:indpol}--\eqref{eq:scmepot} is achieved with
an iterative procedure and a suitable convergence threshold of the 
induced 
moments to achieve energy-force consistency (see the Supplementary Information).

As the point moments come close
the multipole moment expansion breaks down -- resulting 
in the so-called polarization catastrophe.\cite{thole:1981} In order
to avoid this screened interaction tensors are introduced\cite{thole:1981,masia2005,masia2006,burnham1999,Stone:2011,dampingbothyeah}
which effectively smear out
the point moments.
To zeroth order the Coulomb interaction tensors in eq~\ref{eq:scmepot} are defined as
\begin{equation}
    T^{ij} = \frac{1}{|\br^j - \br^i|}\lambda_0(r) = \frac{1}{r}\lambda_0(r)
\end{equation}
where $\lambda_0(r)$ is a short-range electrostatic interaction
screening function.
The gradient operators act to increase the order of the screened
interaction tensors, for example
\begin{align}
     \nabla_\al T^{ij} =& T_\al^{ij} \equiv -\frac{r_\al}{r^3}\lambda_1(r) \\
      \nabla_\be T_\al^{ij} =& T_\ab^{ij} \equiv 
                             3\frac{r_\al r_\be}{r^5}\lambda_2(r) 
                             - \frac{\delta_\ab}{r^3}\lambda_1(r)
\end{align}
where $r_\al = (\br^j - \br^i)_\al$.

Most commonly used interaction tensor screening functions in the
context of polarizable force fields are based on exponential
decay of the point charges resulting in the Thole-type damped 
tensors.\cite{thole:1981} Here we make use of screening functions 
derived from considering the overlap and resulting Coulomb 
electrostatic screening of Gaussian charge densities and 
multipoles.\cite{Stone:2011}
In the
equations above they are
\begin{align}
 \lambda_1(r) =&\ \mr{erf}(S) - \frac{2}{\sqrt\pi}Se^{-S^2} \\
 \lambda_2(r) =&\ \mr{erf}(S) - \frac{2}{\sqrt\pi}\left(S + \frac{2}
 {3}S^3\right)e^{-S^2} \label{eq:damp1}
\end{align}
where $S$ is the screened distance, $S=r/g$, and $g$
is the screening length -- describing the spatial extent of the
Gaussian functions.  

In the SCME/f model the total energy is a functional of the external field, $V^i_\al$, 
at each molecular \ecom\ site $i$ and is given by
\begin{equation}
    E_\mr{tot}[\{V^i_\al\}] = E_\mr{elst}[\{V^i_\al\}] 
                      + E_\mr{non-elst} + E_\mr{mon} \label{eq:etot}
\end{equation}
where the terms on the right hand side are, $E_\mr{elst}[\{V^i_\al\}]$, 
the total electrostatic energy functional,
the non-electrostatic terms, $E_\mr{non-elst}$, which includes a pair-wise repulsive
and a dispersion potential, and $E_\mr{mon}$, 
which is a sum of the internal energies described by the Partridge-Schwenke 
potential energy surface (PS--PES) of the water monomer.\cite{partridge1997}

More explicitly the 
first term on the right hand side of eq~\eqref{eq:etot}
can be further separated into three terms describing the inter- and 
intramolecular contributions to the total
electrostatic energy of the
system, namely
\begin{equation}
 E_\mr{elst}[\{V^i_\al\}] = E_\mr{in}[\{V^i_\al\}] + E_\mr{pol}[\{V^i_\al\}] + E_\mr{self}[\{V^i_\al\}]
\end{equation}
where $E_\mr{in}[\{V^i_\al\}]$ is the electrostatic interaction between the intrinsic molecular moments and $E_\mr{pol}[\{V^i_\al\}]$ is the field-induced polarization energy. 
At self-consistency these terms combine to give
\begin{align}
    E_\mr{in+pol}[\{V^i_\al\}]
    =& \frac{1}{2}\sum^n_i
 \bigg((\mi_\al(\{\br^{ia}\})+\Delta\mi_\al)V^i_\al
 + \frac{1}{3}(\qi_\ab(\{\br^{ia}\}) + \Delta\qi_\ab)V^i_\ab 
  \nonumber \\
 &\ \ \ \ \ \ \ \ \ + \frac{1}{15}\oi_\aby V^i_\aby + \frac{1}{105}\hi_\abyd V^i_\abyd\bigg)
\end{align}
$E_\mr{self}$ is the on-site self-energy, given by 
\begin{equation}
    E_\mr{self}[\{V^i_\al\}] = - \frac{1}{2}\sum_i^n\left(\Delta\mi_\al V^i_\al 
  + \frac{1}{3}\Delta\qi_\ab V^i_\ab\right)
\end{equation}
and accounts for the change in internal energy required to polarize the
molecules.

The non-electrostatic term is composed of two intermolecular pair-wise
potentials centered on the oxygen atom
\begin{equation}
    E_\mr{non-elst} = E_\mr{rep} + E_\mr{disp} 
\end{equation}
describing repulsion, $E_\mr{rep}$, and dispersion $E_\mr{disp}$. In the
following expressions for the potentials the distance $r$ refers to
the oxygen-oxygen distance between pair $i$ and $j$, or 
$r = |\br^{j\mr{O}} - \br^{i\mr{O}}|$.

We make use of the same dispersion coefficients as in the original
SCME model\cite{scme2}. The dispersion energy is
\begin{equation}
    E_\mr{disp} = -\sum_i^n\sum_{j<i}^n\left(
      \frac{C_6}{r^6}t_6(r) 
    + \frac{C_8}{r^8}t_8(r)
    + \frac{C_{10}}{r^{10}}t_{10}(r)
    \right)
\end{equation}
with isotropic coefficients up to tenth order from Wormer and Hettema\cite{wormer1992many}. 
At short range the
interaction is smoothly switched off with a Tang-Toennies damping 
function\cite{tang1984improved}
\begin{equation}
    t_m(r) = 1 - e^{-\tau_dr}\sum_{k=0}^m\frac{(\tau_dr)^k}{k!}
    \label{eq:tt_damping}
\end{equation}
where the parameter $\tau_d$ represents the inverse
decay length of the charge density. 

In the rigid SCME\cite{scme2} model a modified Born-Mayer potential
is used, which includes a term which scales the magnitude of the repulsion
depending on the local environment around the repulsion center --
a molecular density dependent term. With the introduction of the 
Gaussian type 
interaction tensor 
screening function 
we find the molecular density dependence unnecessary and revert back to the
basic Born-Mayer type potential. The pair-wise repulsion is
\begin{equation}
    E_\mr{rep} = \sum_i^n\sum_{j<i}^n A_\mr{rep}r^{-k}e^{-hr}
    \label{eq:repulsion}
\end{equation}
The parameters of the non-electrostatic terms, $\tau_d$, $A_\mr{rep}$, $k$
and $h$, are optimized to work with the new SCME/f model. The
optimization also includes the screening length parameter $g$ of eq~\eqref{eq:damp1}.
The fitting is described in section \ref{sec:model_fit}.

\section{\label{sec:dms_qms} The Dipole and Quadrupole Moment Surfaces}

The internal energy as described by the PS-PES includes
analytical atomic force components,\cite{partridge1997} 
as well as an accurate mapping 
of the dipole moment surface (DMS) for an isolated water 
molecular as a function of the internal geometry. The DMS
is given by
\begin{equation}
    \mi_\al(\br^{i\mr{O}},\br^{i\mr{H}_1},\br^{i\mr{H}_2}) 
    =  q^{i\mr{H}_1}r^{i\mr{H}_1}_\al 
      +q^{i\mr{H}_2}r^{i\mr{H}_2}_\al
      +q^{i\mr{O}}r^{i\mr{O}}_\al \label{eq:dms}
\end{equation}
where $q^{i\mr{O}} = -(q^{i\mr{H}_1} + q^{i\mr{H}_2})$ and the partial charges of the two hydrogens are in turn a function
of the internal geometry, fitted to recreate the calculated DMS.
For example 
$q^{i\mr{H}_1} = q^{i\mr{H}_1}(r^{\mr{OH_1}},r^{\mr{OH_2}},
\mr{cos}(\theta_\mr{HOH}))$, where $r^{\mr{OH_1}}$ and 
$r^{\mr{OH_2}}$ are the internal bond lengths between the oxygen and 
the two hydrogens, and $\theta_\mr{HOH}$ the HOH angle. We make use 
of this mapping, and leave it unchanged. 

\begin{figure}[!th]
\includegraphics[width=.35\textwidth]{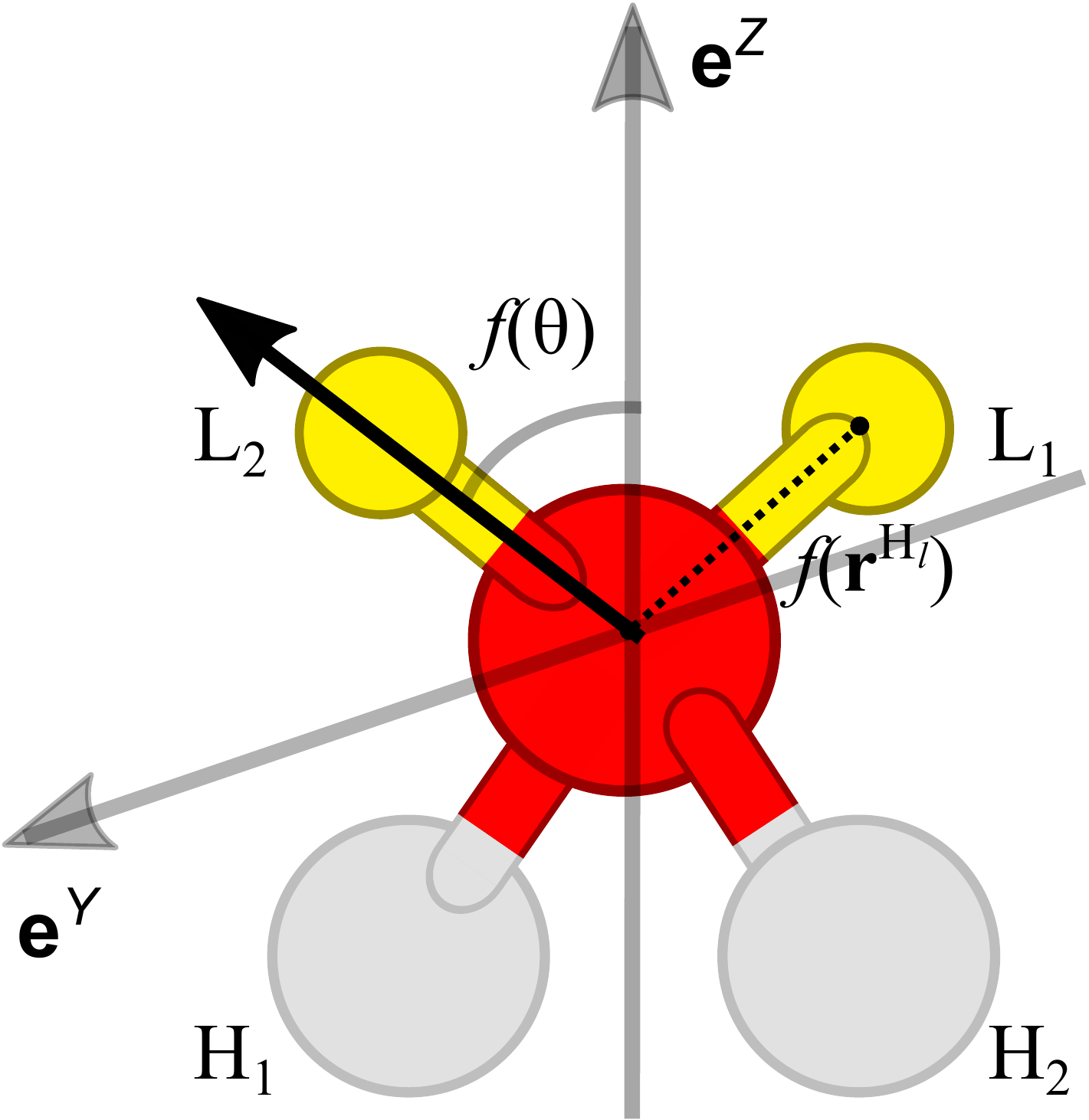}  
\caption{%
L-site placement (yellow) in the water monomer structure. 
The relationship of the angle to the unit basis vectors which describe the local reference frame is shown, eq~\eqref{eq:geometry-theta} and eqs~\eqref{eq:rot1}--\eqref{eq:rot2}. 
For example, operating with the rotation vector corresponding to hydrogen indexed 1 on $e^{iZ}_\al$ results in $\left(\mr{cos}(f(\theta))e^{iZ}_\al - \mr{sin}(f(\theta))e^{iY}_\al\right)$. 
Due to symmetry specific indexing of the atoms is completely interchangeable, and either pair of H and L in the Figure above can serve as pair 1 or 2.
The distance from the oxygen to a L-site, controlled with $f(\br^{\mr{H}_l})$ is a second order polynomial function depending on the position of one of the hydrogens (while the position of the other L-site depends on the other hydrogen), eq~\eqref{eq:geometry-length}.
Positions and scales are exaggerated for clarity.
}
\label{fgr:l-sites}
\end{figure}

The DMS partial charges are not suitable to 
describe a quadrupole moment surface (QMS) without modification. Instead 
the charge site associated with the oxygen is split up into 
two components and placed within a plane perpendicular to the 
symmetry plane of the hydrogens and oxygens. The sites 
are denoted L$_1$ and L$_2$, where the site positions are 
directly related to the length of the hydrogen bond lengths 
indexed H$_1$ and H$_2$, and the HOH angle. See Figure \ref{fgr:l-sites}. 
The QMS is
written as
\begin{equation}
    \qi_\ab(\br^{i\mr{O}},\br^{i\mr{H}_1},\br^{i\mr{H}_2}) = \sum_{a}
    ^{n'_i} 
    \frac{3}{2}\bigg\{q^{ia}\left(
    (\br^{ia}-\br^i)_\al(\br^{ia}-\br^i)_\be
    -\frac{\delta_\ab}{3}||\br^{ia}-\br^i||\right)\bigg\} \label{eq:QMS}
\end{equation}
where $n'_i$ denotes the sites 
$\{\mr{H}_1',\mr{H}_2',\mr{L}_1,\mr{L}_2\}$ associated with molecule $i$.
The apostrophe on the hydrogen is used to discern their role in the QMS from
their role in the DMS since
the charges $q^{i{\mr{H}_l}'}$ are different from the DMS 
charges, and are
\begin{equation}
    q^{i{\mr{H}_l}'} = \mr{A}q^{i{\mr{H}_l}} + \mr{B}q^{\mr{H}}_\mr{eq}
    \label{eq:effchrg}
\end{equation}
and for the L-sites they are
\begin{equation}
    q^{i\mr{L}_l} = \mr{C}q^{i\mr{H}_l} + \mr{D}q^{\mr{H}}_\mr{eq}
    \label{eq:effchrgL}
\end{equation}
where $q_\mr{eq}^\mr{H}$ is the DMS charge of the hydrogen in the
equilibrium monomer configuration. 

The position of the L$_1$ and L$_2$ charge sites is related 
to the atomic positions of each water molecule
through a rotation operator times a scaling factor which
controls the length of the 
rotated vector. A translation
operator translates the vector to the \ecom\ position of molecular
site $i$ for completeness. Explicitly this operation is
\begin{equation}
    r^{i\mr{L}_l}_\al = R^{i\mr{L}_l}_{\eta\al}e^{iZ}_\eta f(\br^{\mr{H}_l}) + r^{i}_\al
\end{equation}

We make use of the 
unit basis vectors previously used to define the local-to-global 
rotation matrices in eqs \eqref{eq:com}--\eqref{eq:cross}. The
rotation matrices for the L$_1$ and L$_2$ sites are
\begin{align}
    {\bf R}^{i\mr{L}_1} =& \left( \mr{cos}(f(\theta)){\bf I} -
    \mr{sin}(f(\theta))\left[{\bf e}^{iX}\right]_\times\right) \label{eq:rot1} \\
    {\bf R}^{i\mr{L}_2} =& \left( \mr{cos}(f(\theta)){\bf I} +
    \mr{sin}(f(\theta))\left[{\bf e}^{iX}\right]_\times\right) \label{eq:rot2}
\end{align}
and is a simplification of the general Rodrigues' rotation operator\cite{rodrigues}
in terms of the local orthonormal basis vectors (shown 
in Figure \ref{fgr:localframe}).

In order to allow for flexibility of the L-sites and correlate 
their positions to the change in the positions of the hydrogens, both the angle
factor
and length scale factor are defined in terms of the OH bond lengths and
HOH angle through
\begin{align}
    f(\br^{\mr{H}_l}) =& \mr{a} + \mr{b}(|\br^{i\mr{O}}-\br^{i\mr{H}_l}| -
    r_\mr{eq}) + \mr{c}(|\br^{i\mr{O}}-\br^{i\mr{H}_l}| -
    r_\mr{eq})^2 \label{eq:geometry-length} \\
    f(\theta) =& \mr{d} + \mr{e}(\theta - \theta_\mr{eq}) \label{eq:geometry-theta}
\end{align}
where $r_\mr{eq}$ and $\theta_\mr{eq}$ are the equilibrium hydrogen 
to oxygen bond length and HOH angle of the isolated PS--PES water molecule,
respectively, see Figure \ref{fgr:l-sites}. 
We find that a second order polynomial in terms of the change in bond length, and a linear term for the change in bond angles is adequate
to capture the QMS with good accuracy. 
The charge scaling parameters A, B, C and D, and the geometric parameters a, b, c, d, e
are fitted parameters, described below.

\subsection{Ab initio QMS Calculations and Fit}

The dipole and quadrupole moment is mapped using
the ab initio quantum chemistry software ORCA\cite{ORCA1,ORCA2}.
An iterative-configuration expansion configuration interaction (ICE-CI)
method is used, with the aug-cc-pvqz basis set and the 
energy convergence threshold is set to $10^{-8}$ E$_h$. 
Eight correlated electrons are included and the 
active orbitals were chosen by including MP2 orbitals of natural orbital occupation numbers ranging between 1.99999 and 0.00001.
The ICE-CI method is related
to the CIPSI technique.\cite{cipsi} Note that this level of theory is necessary
to accurately determine the dipole and quadrupole moment using their well defined
charge density based operators, instead of resorting to energy based 
schemes to estimate these quantities. For example, we find that coupled-cluster at the
CCSD(T)/aug-cc-pvqz level of theory and
orbital optimized coupled-cluster theory OOCCSD(T)/aug-cc-pvdz, did not provide 
a satisfactory agreement with the DMS of the PS-PES, when using the dipole moment 
operator $\mu_\al = \int\rho(\br)r_\al d\br$. See the Supporting Information for more 
details.

\begin{figure}
\includegraphics[width=.5\textwidth]{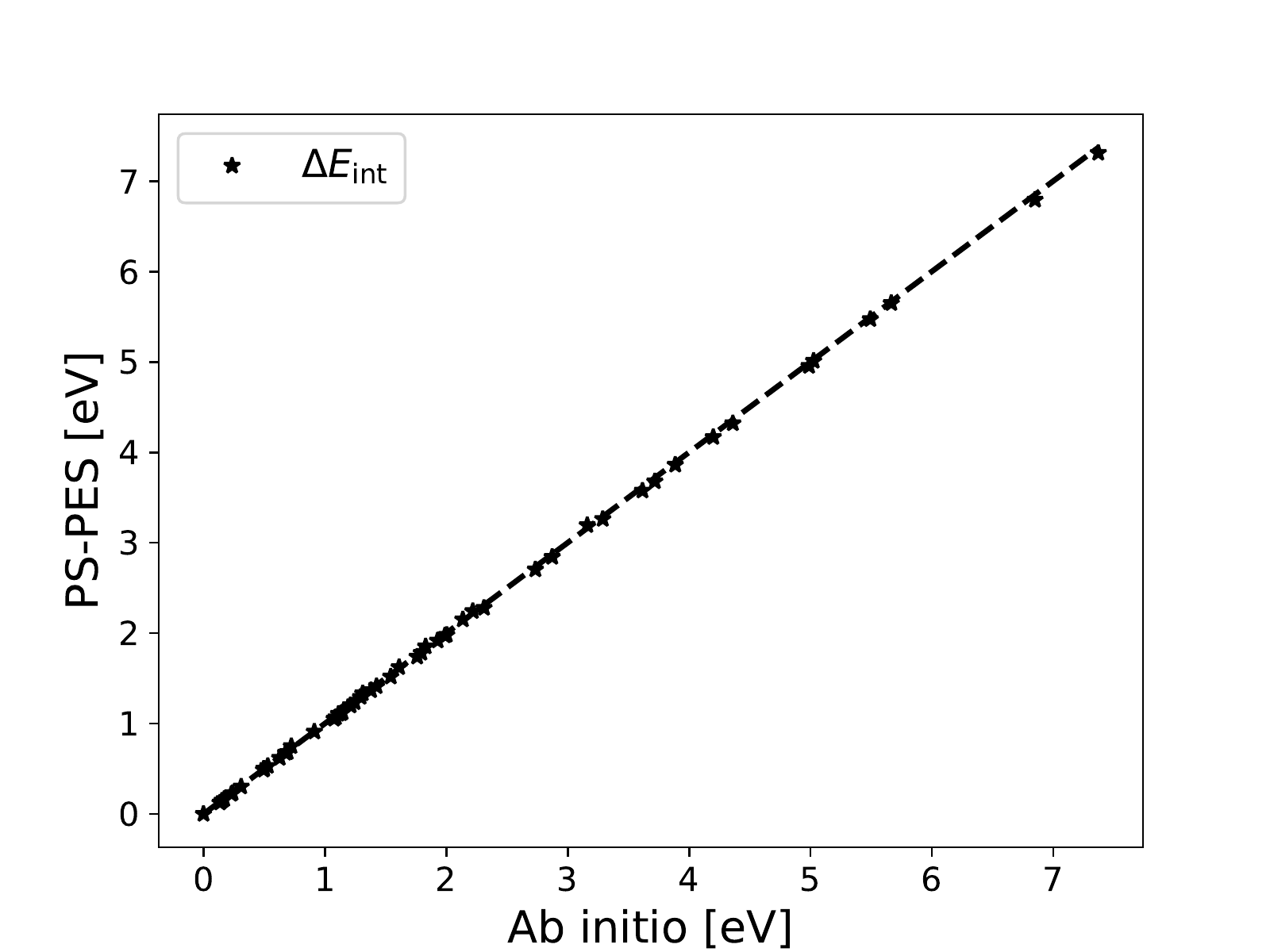}
\caption{%
The relative internal energy difference between the different monomer configurations
used in the QMS fit, compared between the ab initio results and the PS--PES. The 
good agreement between the two methods implies that the use of the ICE-CI data to fit 
the QMS justifies the use of the original PS--PES to
represent internal energy changes and resulting atomic forces, as both potential energy
surfaces are close with RMSD of 0.022 eV, within chemical 
accuracy ($\sim$0.51 kcal/mol).
}
\label{fig:deltaE}
\end{figure}

\begin{figure}[!th]
\includegraphics[width=.49\textwidth]{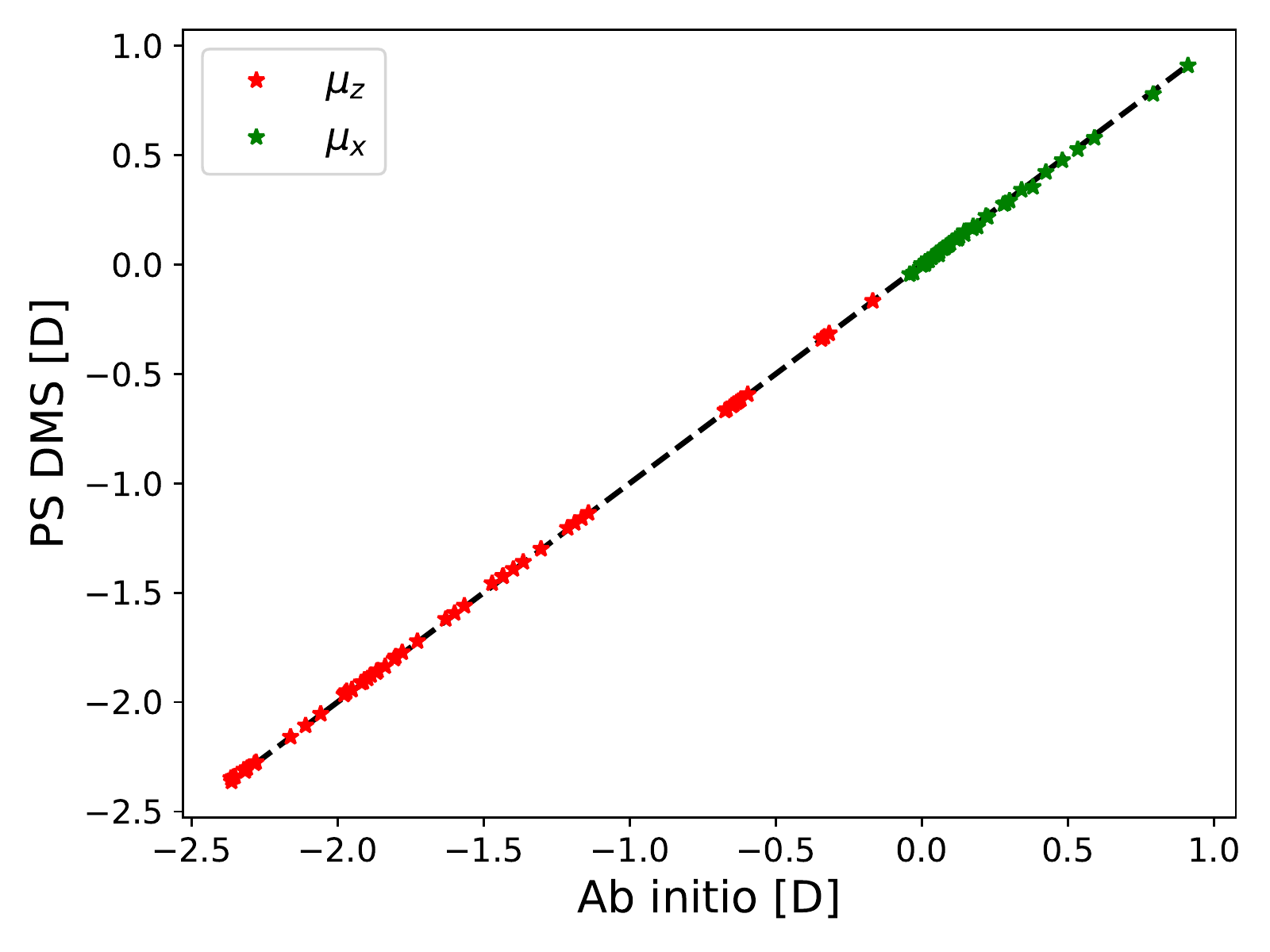} 
\includegraphics[width=.49\textwidth]{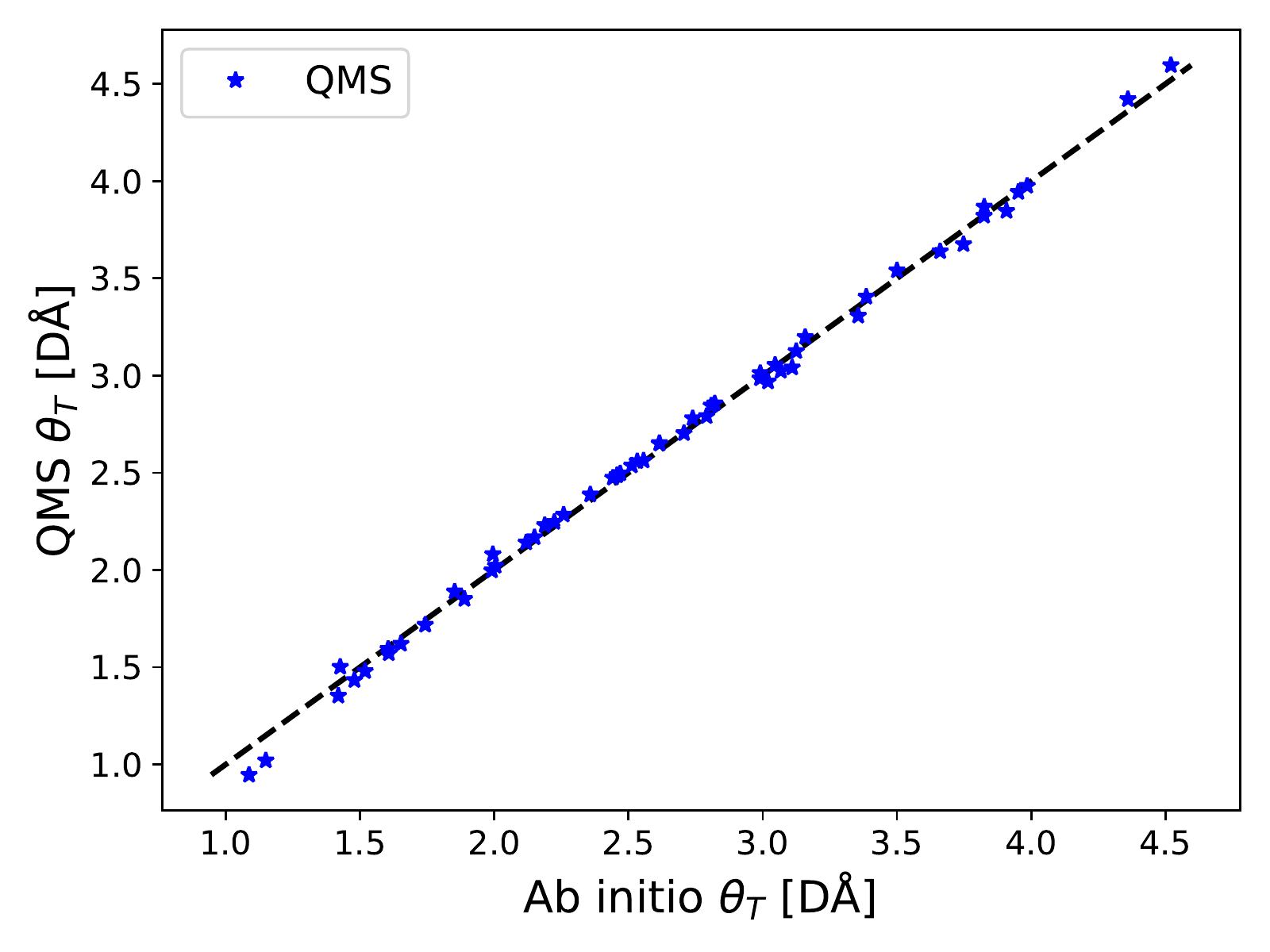}
\caption{%
Left: comparison of the dipole z- and x-components, $\mu_z$ and $\mu_x$ respectively, 
as predicted by the DMS, 
eq~\eqref{eq:dms} and compared to the ICE-CI $\mu_z$ and $\mu_x$. Note that
due to a choice of local reference frame the $\mu_y$ component is always numerically zero.
The DMS of the PS-PES and ICE-CI are in an excellent agreement, with a RMSD of 0.004 D and within 0.5\% on average.
Right: comparison of the $\theta_T$ component mapped by the QMS, eq~\eqref{eq:QMS}, 
with the ab initio ICE-CI data. 
The geometric QMS model of this work, which is fitted to best reproduce the
ab initio results, captures the results to a good degree with low scatter, a mean absolute error of 0.04 D\AA, and 
an average RMS difference of around 1.6$\%$ (see Supplementary Information for the RMSD analysis).
} 
\label{fig:DMSQMS}
\end{figure}

Starting from the ground state geometry in the local-frame as shown in Figure \ref{fgr:localframe} the internal bond lengths and HOH angle are
systematically changed and range from 0.7-1.3 \AA, and 60-175$^{\circ}$, respectively.
These intervals broadly represent the variation in the bond lengths and the angle of the water molecule
in the liquid phase at ambient conditions. Figure \ref{fig:deltaE} shows a comparison 
between the internal energy change of each configuration as calculated by the 
ICE-CI method compared to the PS-PES. The agreement is excellent, and justifies 
the use of the ab initio data to fit the QMS while retaining the original PS-PES 
energy mapping to describe the internal energy change and resulting
atomic forces in our model. Figure \ref{fig:DMSQMS}, left, presents a comparison
between the ICE-CI DMS and the PS-PES DMS, again in an excellent agreement. 

The QMS model parameters associated with the charges in eqs~\eqref{eq:effchrg}--\eqref{eq:effchrgL}, A, B, C and D, as well as the geometric parameters of 
eqs~\eqref{eq:geometry-length}--\eqref{eq:geometry-theta}, a, b, c, d and e, 
are fitted to best reproduce the principal quadrupole moment component. Considering
the water molecule in the ground state configuration the symmetric quadrupole moment
tensor can be written as
\begin{equation}
    \theta = \begin{bmatrix}
             \theta_{T}-\Delta & 0 & 0 \\
             0 & -\theta_{T}-\Delta & 0 \\
             0 & 0 & 2\Delta
            \end{bmatrix} \label{eq:theta}
\end{equation}
where $\theta_T = (\theta_{xx} - \theta_{yy}) / 2$. 

The values of the QMS parameters are determined by carrying out a 
least-squares
optimization, using a module 
freely available in the scientific computing package SciPy.\cite{2020SciPy-NMeth}
Table \ref{tbl:qms-para} presents the numerical values and units of 
the resulting best fit parameters, and Figure \ref{fig:DMSQMS}, right,
shows the resulting fit of the $\theta_T$ components, compared between
the QMS fit and ab initio ICE-CI values. The overall fit is in good agreement
with the ab initio values over a broad range of $\theta_T$ values, with very low
scatter. The largest deviation is found where $\theta_T$ is lowest, i.e. where the quadrupole moment interaction strength is the weakest.

\section{\label{subsec: ana_for} Forces}

With the various expressions given in the preceding section 
analytical atomic force components can be obtained and are 
derived from 
the negative gradient of the total energy expression, eq~\eqref{eq:etot}, 
with respect to the position of atom $a$ in
molecule $i$, or
\begin{align}
    F^{ia}_\al =& -\frac{dE_\mr{tot}}{dr^{ia}_\al} \nonumber \\
    =& -\frac{\partial E_\mr{elst}}{\partial r^{ia}_\al} 
      -\frac{\partial E_\mr{non-elst}}{\partial r^{ia}_\al}
      -\frac{\partial E_\mr{mon}}{\partial r^{ia}_\al}
      \label{eq:forces1}
\end{align}
The first term on the right hand side result in 
several contributing factors to the atomic forces due 
to the definition of the principal axes, choice of 
expansion center and the DMS and QMS. 
The atomic forces resulting from the simple pair-wise potentials 
describing the
non-electrostatic terms are omitted for the sake of brevity,
and the atomic forces due to the monomer energy expression --
the PS-PES -- are accounted for in their original work.\cite{partridge1997}

The first term on the right hand side of eq~\eqref{eq:forces1}, 
the total intermolecular electrostatic
interaction, can be further divided into four contributions
\begin{align}
    -\frac{\partial E_\mr{elst}}{\partial r^{ia}_\al} =&
    -\frac{\partial E_\mr{elst}}{\partial \mu^j_\be(\{\br^{jb}\})}
     \frac{\partial \mu^j_\be(\{\br^{jb}\})}{\partial r^{ia}_\al} 
    -\frac{\partial E_\mr{elst}}{\partial \theta^j_\by(\{\br^{jb}\})}
     \frac{\partial \theta^j_\by(\{\br^{jb}\})}{\partial r^{ia}_\al}
     \nonumber \\
    &-\frac{\partial E_\mr{elst}}{\partial V^j_{\byde\dots\eta}}
     \frac{\partial V^j_{\byde\dots\eta}}{\partial r^{ia}_\al}
     -\frac{\partial E_\mr{elst}}{\partial R^j_{\eta\be}}
      \frac{\partial R^j_{\eta\be}}{\partial r^{ia}_\al}
\end{align}
which are, in order, the partial derivative of the DMS and
QMS, partial derivative of the external field and gradients thereof, 
and 
partial derivatives of the local-to-global rotation matrices 
as defined in eqs~\eqref{eq:com}--\eqref{eq:rot_mat}. 

At self-consistency of the iterative process which minimizes the energy
in terms of the polarized 
moments the following conditions apply
\begin{equation}
  \frac{\partial E_\mr{elst}}{\partial \dmi_\al} 
= \frac{\partial E_\mr{elst}}{\partial \dqi_\ab}
= \frac{\partial E_\mr{self}}{\partial \dmi_\al}  
= \frac{\partial E_\mr{self}}{\partial \dqi_\ab} 
= 0 \nonumber
\end{equation}
There are no explicit force contributions from the self-energy terms
due to the on-site external field as the self-energy can be written solely
in terms of the on-site induced 
moments (see the Supplementary Information). 
This results in a non-trivial additional condition
\begin{equation}
 \frac{\partial E_\mr{self}}{\partial V^j_{\byde\dots\eta}} = 0
\end{equation}
Due to these conditions 
of the self-energy
a single force contribution arises and is due to the local-to-global 
transformation of the fixed polarizability tensors
\begin{equation}
    -\frac{\partial E_\mr{self}}{\partial r^{ia}_\al} = 
    -\frac{\partial E_\mr{self}}{\partial R^j_{\eta\be}}
     \frac{\partial R^j_{\eta\be}}{\partial r^{ia}_\al}
\end{equation}
The total force contribution due to the intermolecular electrostatic
interaction and intramolecular self-energy is then
\begin{align}
    -\left(\frac{\partial E_\mr{in+pol}}{\partial r^{ia}_\al}+
    \frac{\partial E_\mr{self}}{\partial r^{ia}_\al}\right)=&
    -\frac{\partial E_\mr{in+pol}}{\partial \mu^j_\be(\{\br^{jb}\})}
     \frac{\partial \mu^j_\be(\{\br^{jb}\})}{\partial r^{ia}_\al} 
    -\frac{\partial E_\mr{in+pol}}{\partial \theta^j_\by(\{\br^{jb}\})}
     \frac{\partial \theta^j_\by(\{\br^{jb}\})}{\partial r^{ia}_\al}
     \nonumber \\
    &-\frac{\partial E_\mr{in+pol}}{\partial V^j_{\byde\dots\eta}}
     \frac{\partial V^j_{\byde\dots\eta}}{\partial r^{ia}_\al}
     -\left(\frac{\partial E_\mr{in+pol}}{\partial R^i_{\eta\be}}
      +\frac{\partial E_\mr{self}}{\partial R^i_{\eta\be}}\right)
      \frac{\partial R^i_{\eta\be}}{\partial r^{ia}_\al}
      \label{eq:force-terms}
\end{align}
The terms in the expression above are given explicitly in the 
Supporting Information. We note that in order to evaluate the 
first term on the right hand side, explicit partial charge 
derivatives with respect to atomic positions of the DMS are required, 
which were not included in
the original work on the PS--PES.\cite{partridge1997} These are provided 
by Burnham and Xantheas, first used in the development of a 
flexible Thole-type multipole moment expansion potential.\cite{burnham2002development}


\section{\label{sec:model_fit} Flexible Model Fit}

With the introduction of the DMS and QMS, 
the Gaussian type interaction tensor screening
functions, as well as the changes to the pair-wise repulsion function, 
all of the five model parameters which affect the intermolecular interactions $g$, $\tau_d$, $A_\mr{ref}$, $k$ and $h$ are 
re-fitted. The fitting is performed with the same 
least-squares optimization module used for the QMS fit. We make use of the same numerical values 
for the fixed octupole and hexadecapole, as well as the dipole-dipole, dipole-quadrupole and 
quadrupole-quadrupole polarizability as in the original SCME model.\cite{scme2} 
The Fortran based SCME/f code
is freely available online\cite{UnifiedSCME}, and includes an interface to
the Python based Atomic Simulation Environment\cite{ASE,ASE2017} library.

\begin{figure}
\includegraphics[width=0.9\textwidth]{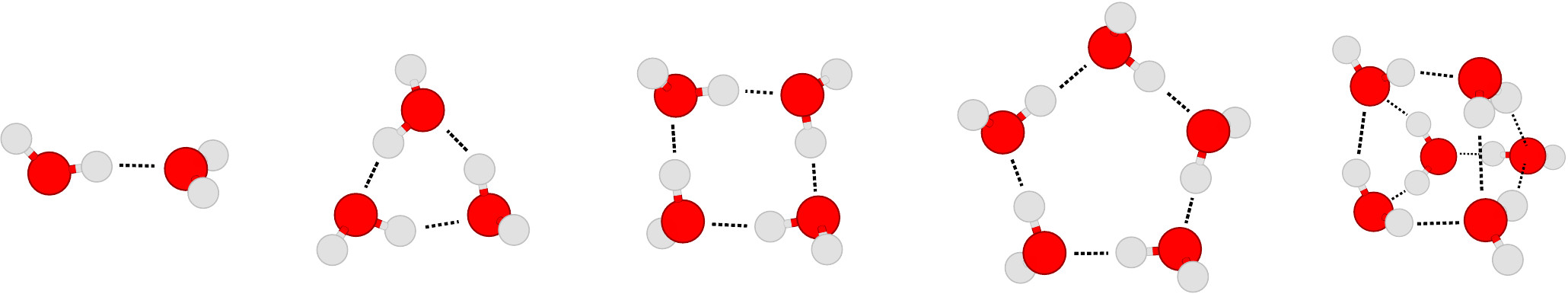}
\caption{The lowest lying water cluster $(\ce{H2O})_n$ isomers for $n$=2-6 used 
in the fitting procedure for the model parameters. From left to right; dimer (Cs),
trimer (UUD), quadromer (S4), pentamer (cyclic, CYC) and hexamer (prism, PRI).}
\label{fig:small_clusters}
\end{figure}

The data set used for the fitting includes several points around the 
minimum of the dimer binding curve with the energy minimum and oxygen-oxygen distance
corresponding to
CCSD(T) calculations.\cite{temelso:2011}
A single interaction energy for the lowest lying trimer, tetramer,
pentamer and hexamer is included. 
Figure \ref{fig:small_clusters} shows the geometry of the 
lowest-lying water clusters
$(\ce{H2O})_n$ in the range $n=2-6$.
The reference calculations which we make use of here 
include the complete basis set limit CCSD(T)
energies of the low-lying water hexamer structures by Bates and Tschumper.\cite{bates:2009} For the other cluster sizes -- trimers, tetramers 
and pentamers -- complete basis set limit RI-MP2 calculations, 
with CCSD(T) corrections, are used.\cite{temelso:2011}

In addition to the clusters we have also considered properties of hexagonal ice (ice Ih), which is the most common ice phase.
There are no high-level first-principles calculations with sufficient accuracy to serve as reference values.
Instead, we need to use experimental data for lattice constants, unit cell volume, bulk modulus and lattice energies which generally include zero-point energy (ZPE) effects, and these effects are quite sizeable.\cite{whalley:1957,whalley:1958,whalley:1984,rasti2019importance}
Consequently, we have performed phonon calculations with the SCME/f model for 
proton disordered
units cells of ice Ih containing 96 water molecules using the Parlinski-Li-Kawazoe finite-displacement method \cite{parlinski1997} 
as implemented in the \textsc{phonopy} package\cite{togo2015} using $3 \times 3 \times 3$ supercells and a displacement of 0.01 \AA.
For a unit cell with fixed cell vectors we have first performed a geometry relaxation employing the analytical SCME/f forces with a force threshold of $10^{-3}$ eV/\AA.
Then, using a $10 \times 10 \times 10$ q-point sampling for the Brillouin zone integration, 
we obtain a numerically converged phonon density of states $g(\omega)$, the first moment of which provides the ZPE
\begin{equation} 
\label{eq:ZPE}
    E_\mr{ZPE} = \tfrac{\hbar}{2} \int_0^\infty \omega \, g(\omega) \, d\omega \quad .
\end{equation}
Considering the dependence of the phonon frequencies on the unit cell volume $\omega = \omega(V)$
within the so-called quasi-harmonic approximation yields a ZPE-corrected energy-volume curve
\begin{equation} 
  E_\mr{tot+ZPE}(V) = E_\mr{tot}(V) + E_\mr{ZPE}(V)
  \quad ,
\end{equation}
where the energy zero is such that it describes infinitely separated (non-bound) individual water molecules. 
By fitting the Rose-Vinet equation of state \cite{vinet1987} we obtain the minimum $E_\mr{lat}^\mr{ZPE} = E_\mr{tot+ZPE}(V_0^\mr{ZPE})$ of that curve together with the ZPE-corrected bulk modulus $B_0^\mr{ZPE}$
(see supporting information for more details),
which can be compared against accurate experimental data.\cite{whalley:1957,Rottger:sh0050,hobbs2010ice}
In order to include this data in the fitting process, an initial $E_\text{tot+ZPE}(V)$ was calculated based on the SCME/f parameters first determined by fitting the data set derived from the water clusters.
Then, $E_\text{tot}(V)$ was improved by further parameter adjustments such that the expected ZPE correction would bring it close to the experimental values. 
This trial and error scheme was found necessary since the phonon calculations are significantly more expensive than the calculation of the cluster properties.
The end results based on a new set of phonon calculations is presented in Table~\ref{tbl:ice} and shows good agreement with the experimental target properties.
(The concomitant energy-volume curves are shown in the supporting information.)
Table \ref{tbl:modelpara} compiles the concomitant final optimized parameters of the SCME/f model.

Table \ref{tbl:clusters} shows the resulting interaction energy and relative interaction energy versus the reference CCSD(T) calculations of the lowest lying isomers used in the fit. This includes a structural analysis comparing the relaxed SCME/f structure to the 
CCSD(T) reference structures, where the RMS deviation of nearest neighbor oxygen-oxygen distances , $\braket{d\br_\mr{OO}}$, intramolecular oxygen-hydrogen bond lengths of the donor hydrogens, $\braket{d\br_\mr{OH}}$, hydrogen bonding (H-bond) bond lengths, $\braket{d\br_\mr{O\cdots H}}$, and 
angles 
between oxygen-hydrogen-oxygen in H-bonds, $\braket{d\angle \mr{OHO}}$, are presented. The overall RMSD
of the atomic positions is also presented, $\braket{d\br^a}$, and is evaluated with the Kabsch algorithm.\cite{kabsch:1976} The interaction energies for the different cluster sizes are reproduced
to a reasonable degree, with sub kcal/mol difference compared to the CCSD(T) results, except for 
the prism isomer of the hexamer where the interaction energy is overestimated by $1.18$ kcal/mol. 
The resulting relaxed structures are in an overall very good agreement with the reference structures,
with small variations in the second decimal in terms of atomic distances.  Similarly, the angles 
between OHO in H-bonds are in a good agreement with the reference. The largest deviation is found in the angle between the donor-acceptor in the dimer.

\section{\label{sec:validation} Model Validation}

\begin{figure}[!th]
\includegraphics[width=.75\textwidth]{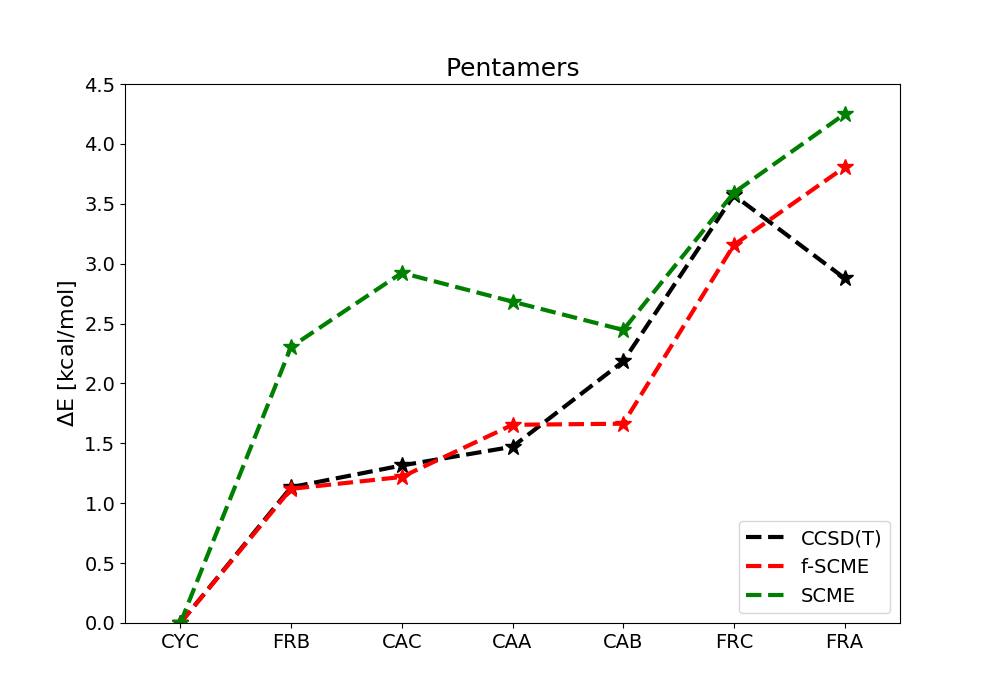}
\includegraphics[width=.75\textwidth]{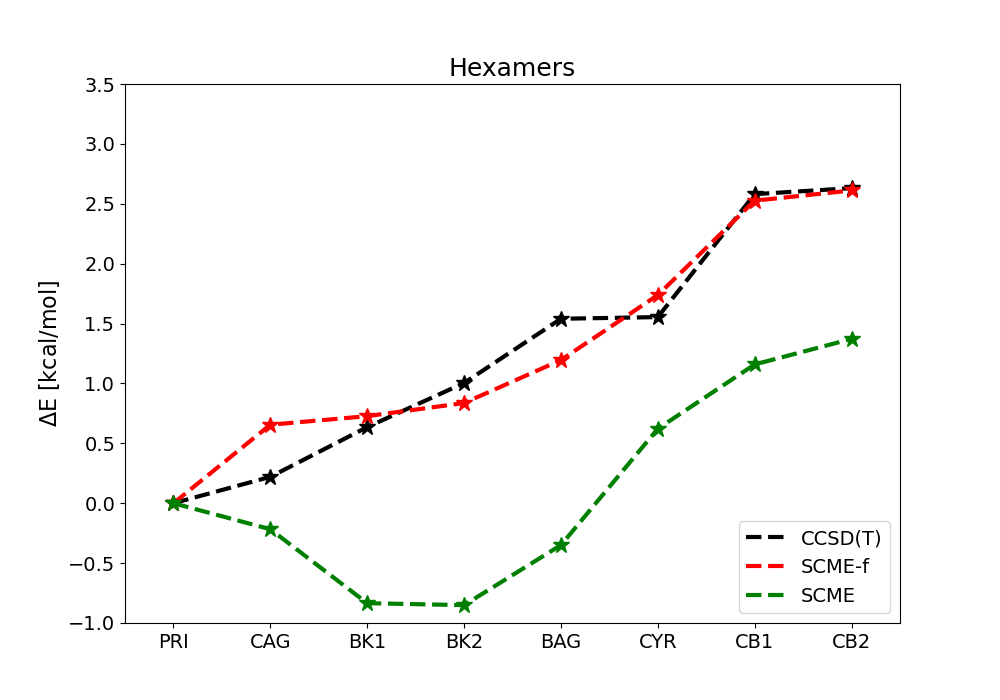}
\caption{Relative energy difference for the lowest lying pentamers (top) and
hexamer (bottom) water cluster isomers. 
The results for the rigid version of SCME\cite{scme,scme2}
and SCME/f are compared.
Relative energy differences from high level quantum
chemistry calculations are also shown; for the pentamers these are RI-MP2 energies at
the complete basis set limit with CCSD(T) corrections\cite{temelso:2011} 
(MP2/CBS+$\Delta$CCSD(T)); for the hexamers these are CCSD(T) energies at 
the complete basis set limit (CCSD(T)/CBS).\cite{bates:2009} The acronyms from left to right 
are the different isomers. Pentamers; cyclic (CYC), fused-ring-B (FRB), cage-C
(CAC), cage-A (CAA), cage-B (CAB), fused-ring-C (FAC) and fused-ring-A (FRA); and the 
hexamers; prism (PRI), cage (CAG), book-1 (BK1), book-2 (BK2),
bag (BAG), cyclic-ring (CYR), cyclic-boat-1 (CB1) and cyclic-boat-2 (CB2).}
\label{fig:penta-hexa}
\end{figure}

For further validation of the new 
model 
the interaction energies and relative energy differences of all higher lying isomers of the pentamers and hexamers are calculated, which are not included in the fitting data set, and compared to the relative energies from the quantum chemistry references.\cite{bates:2009,temelso:2011} 
The trends are shown in Figure \ref{fig:penta-hexa}, and the trend predicted with the rigid SCME is shown for comparison. 
All structures are relaxed with a force tolerance of 1.0e$-4$ $e$V/\AA, and results collected in Table~\ref{tbl:pentamers-hexamers}, which also presents the RMS difference between the relaxed SCME/f structures and the quantum chemistry reference structures. 

For the pentamers, Figure~\ref{fig:penta-hexa} top, most of the relative energy difference trend is captured with the exception of isomer FRA, whose relative stability is underestimated. 
Another key difference between SCME/f and the reference calculations is the series of CAA-CAB isomers, which have a cagelike structure.
In particular the cage structure of isomers CAA and CAB are not stable and rearrange to isomers which are more akin to the fused ring structures of the FRA-FRC isomers.
The resulting SCME/f structures of CAA and CAB are near identical, with an interaction energy difference of only 0.01 kcal/mol.
Only the CAC isomer keeps its cagelike structure, but one of the H-bonds is not stable (between a donor acceptor oxygen with distance greater than 3.0 \AA), resulting in a rotation of one of the water monomers.
Compared to the rigid SCME predecessor this represents an improvement, in particular for the FRB, CAC and CAA isomers, whose stability is greatly underestimated relative to the CYC isomer.

For the hexamers , Figure~\ref{fig:penta-hexa} bottom, the overall trend in the relative interaction energies is captured to a good degree compared to the CCSD(T) reference, and is a substantial improvement over the rigid SCME model, which greatly underestimates the stability of the prism isomer relative to all other isomers. 
The bond lengths and angles of the hexamer isomers are all in very good agreement with the reference structures, with small differences in the second or third decimal in terms of the bond lengths, and the H-bonded OHO angles deviate by only 2-4$^{\circ}$.  
Table \ref{tbl:vibrations} presents vibrational frequency analysis of the 
lowest lying isomers, including the cyclic ring isomer of the hexamer.
The RMS deviation from near-CBS CCSD(T) calculations\cite{howardvib1,howardvib2} are presented for the different classes of modes. These are intermolecular low-energy vibrarional modes (10-1000 cm$^{-1}$), intramonomer bending modes (1600-1800 cm$^{-1}$) and H-bonded and non-bonded OH stretching modes (ca. 3200-3900 cm$^{-1}$).
For comparison, the same analysis is performed for the SCME/f model, but with the quadrupole moment fixed and corresponding to the numerical value of the quadrupole moment for the ground state monomer configuration.

With the inclusion of the QMS (left column Table \ref{tbl:vibrations}) the low-energy vibrational modes and, in particular, 
the bending modes are in a good agreement with the reference 
calculations. The RMS deviation ranges from 18-23 cm$^{-1}$ and 
7-14 cm$^{-1}$ for the two classes of modes, respectively. The
maximum difference in the bending modes does not exceed 
20 cm$^{-1}$ for any of the clusters analyzed.
The red shift of the 
H-bonded OH stretches is, however, not captured by our model, resulting in an overestimation 
of these modes, which becomes systematically larger with cluster size. This is 
due to the underlying monomer potential energy surface, whose limit in terms of hydrogen dissociation is {\ce{OH^{.}} + \ce{H^{.}}}
whereas should be in the condensed phase {\ce{OH^{-}} + \ce{H^{+}}}. The model 
potential does not describe this important change,
and the resulting weakening of oxygen-hydrogen bonds in H-bonding OH. The high-frequency 
modes for the dimer are though in a reasonable agreement with the reference 
calculations. 

A comparison to the same vibrational frequency analysis is performed with the 
quadrupole moment fixed (right column, Table \ref{tbl:vibrations}). Fixing the quadrupole moment results in a drastic change in the difference between all of the types of modes 
and for all cluster sizes, with for example a RMS deviation of up to three times greater for
the bending modes. The overall agreement with the reference calculations of all modes is consistently worse, in particular for the larger cluster, n=4-6.
Only the low-frequency modes of the Cs dimer seem improved by fixing the quadrupole moment. 
While the parametrization of the intermolecular interaction parameters is 
with the QMS included, the structural properties and interaction energy of 
the small clusters are not drastically changed with the quadrupole moment fixed (see 
Supplementary Information). 

It is also of interest to analyze the structure of the monomers in crystal ice Ih with or without the QMS included. Table \ref{tbl:angles} presents the average internal HOH angle of each water monomer in the crystal lattice, extracted at volume $V^\mr{ZPE}_0$, and compares to the experimental value of the angle for the isolated monomer and in crystal ice Ih. The experiments show a clear widening of the monomer HOH angle by about 3.5 degrees (104.5$^{\circ}$--108.1$^{\circ}$) going from the gas to crystal phase. Without the QMS the trend is opposite, with the angle favoring lower values by about 4.5 degrees (104.5$^{\circ}$--99.95$^{\circ}$), where the dipole moment is high. The correct trend is captured again with the inclusion of the QMS, with the angle widening by about 2 degrees (104.5$^{\circ}$--106.51$^{\circ}$). The QMS correctly balances the magnitude of the dipole moment and principal quadrupole moment in the lattice, and in such a way that a widening of the angle is favoured.

\section{\label{sec:discussion} Discussion and Conclusions}

We have presented an extension of the SCME potential function for water molecules to allow for distortion of the molecular structure. 
In addition to the dipole moment surface, this flexible potential function, SCME/f, includes a mapping of the quadrupole moment surface which has not been previously
included at this level of detail to our knowledge. 
A simpler model for the quadrupole moment that has been used in
both rigid and flexible point charge based potential functions\cite{TIP4P,scott2009,horn2004development,abascal2005potential,abascal2005general},
as well as more sophisticated polarizable models\cite{burnham2002development,ttmn2fr,fanourgakis2008development,burnham2008vibrational,hbbpola,hbbpolb,mbbpola,mbbpolb,mbbpolc}, make use of the so-called M-site.
We now digress in a brief comparison between the QMS model described in this work and the M-site model.

In the M-site model the partial charge associated with the oxygen is moved off the atomic center to a position behind the 
oxygen and on to the bisector defined by the two OH bond vectors. The position of the M-site 
in the global coordinate frame is written as\cite{reimers1982intermolecular,reimers1984structure,suhm1991parameterized}
\begin{equation}
    \br_\mr{M} = (1-\gamma)\br_\mr{O} + \frac{\gamma}{2}(\br_{\mr{H}_1} + \br_{\mr{H}_2}) \label{eq:gammapos}
\end{equation}
where $0 < \gamma \leq 1$. For any finite value of $\gamma$ the partial charges 
are re-scaled according to 
\begin{equation}
    q^{\mr{H}_l^{\gamma}} = \frac{q^{\mr{H}_l}}{1-\gamma},\ \ \ \ \
    q^\mr{M} = -q^{\mr{H}_1^{\gamma}} - q^{\mr{H}_2^\gamma} \label{eq:mch}
\end{equation}
such that the dipole moment remains unchanged 
in the M-site frame, and a single set of three partial 
charges describes both the dipole and quadrupole moment.

More importantly, a value of $\gamma$ can be derived 
such that the $\Delta$ component in eq~\eqref{eq:theta} vanishes, resulting 
in the compactly written moment tensor
\begin{equation}
    \theta = \begin{bmatrix}
             \theta_{T} & 0 & 0 \\
             0 & -\theta_{T} & 0 \\
             0 & 0 & 0 .
            \end{bmatrix} 
    \label{eq:theta_tr}
\end{equation}
This illustrates that the principal quadrupole moment component $\theta_T$ is origin
independent, and is the rational for placing the partial charge on the M-site and not on the oxygen center. The strength of the quadrupole moment interaction is determined by $\theta_T$. For the ground state PE-PES water monomer configuration used in this work a 
$\gamma=0.4071$ results in a compact tensor of the form
in eq~\eqref{eq:theta_tr} (see the Supplementary Information). 
Similar values for $\gamma$ are reported in potential functions based on the M-site. 
While such a three site partial charge model can capture both the
dipole and principal quadrupole moment for 
a fixed ground state monomer configuration, the question is how the model holds up in the case of a flexible water monomer.

\begin{figure}
\includegraphics[width=.49\textwidth]{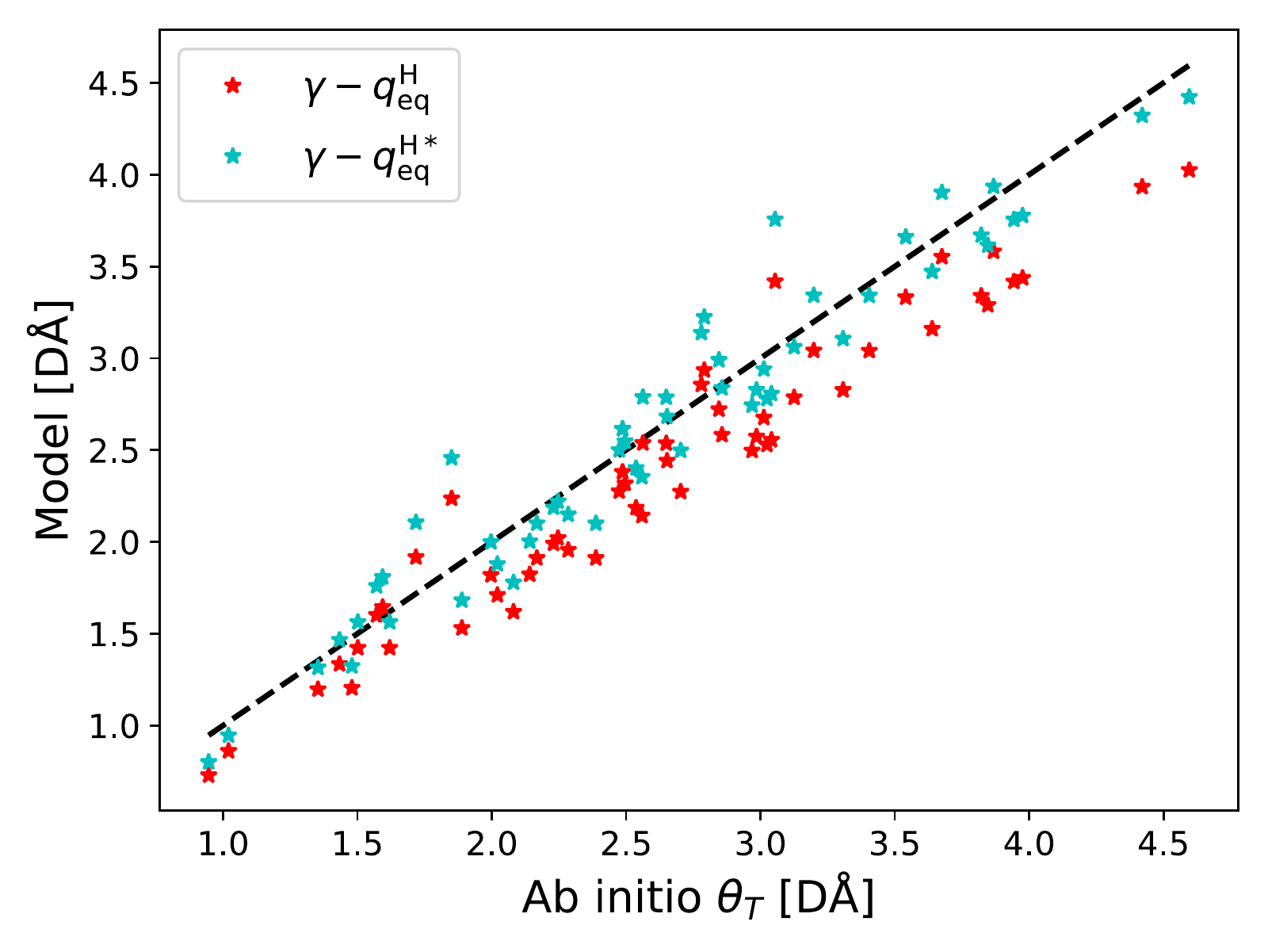} 
\includegraphics[width=.49\textwidth]{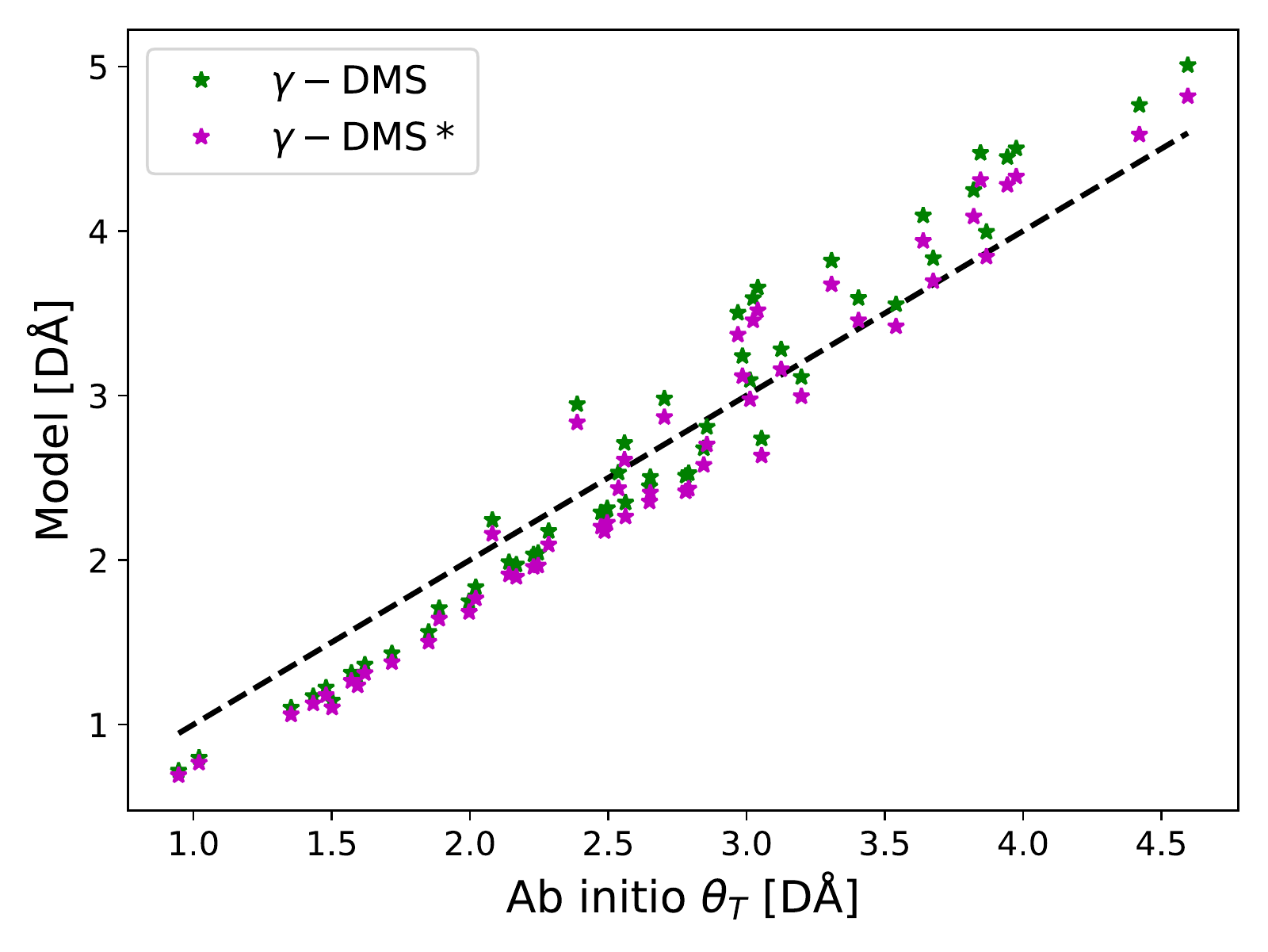}
\caption{%
M-site models. See the Supplementary Information for details on the individual models.
Left: M-site models making use of a fixed charge ($\gamma-q^\mr{H}_\mr{eq}$, red) corresponding to the ground state monomer configuration or scaled charge ($\gamma-q^\mr{H,*}_\mr{eq}$, cyan). 
The fixed point charge model (red)
tends to underestimate the strength of the quadrupole moment over the whole range, 
whereas the scaling results in a change in the slope and an overall
better agreement. 
Right: M-site models making use of variable DMS charges ($\gamma-\mr{DMS}$, green)
and scaled DMS charges 
($\gamma-\mr{DMS}^*)$, magenta).
Both model tend to underestimate the strength of the quadrupole moment
in region of low strength, whereas overestimate in the region of large
strength. The agreement is only slightly improved with the scaled DMS model, corresponding to a shift of the M-site to $\gamma=0.3838$.
}
\label{fig:Mmodels}
\end{figure}

Using the ab inito ICE-CI quadrupole moment data four M-site models are 
considered and compared, and are representative of M-site models encountered in the literature. The details of the models and parameters are presented in the Supplementary Information. 
The first two models, Figure \ref{fig:Mmodels} left, make use of $\gamma=0.4071$ 
and a set of fixed partial charges ($\gamma-q^\mr{H}_\mr{eq}$) -- corresponding to the partial charges of the ground state
monomer configuration -- or scaled ground state charges 
($\gamma-q^\mr{H,*}_\mr{eq}$). The scaling parameter is fit such that the model best captures $\theta_T$ over the whole range. 
The fixed point charge model tends to underestimate the strength of the quadrupole moment
over the whole range, whereas the scaling of the charge results in a change in the slope and overall better agreement. However, in both cases the scatter is substantial and the RMS difference between the trace components of the quadrupole moment versus the ab inito values is $>10$\% on average (see the Supplementary Information). 

In the third and fourth model, Figure \ref{fig:Mmodels} right, the charge are described with the DMS charge. In the third model the optimal $\gamma$ value is used ($\gamma-\mr{DMS}$) and in the fourth model the DMS charges are scaled ($\gamma-\mr{DMS}^*$) to best capture $\theta_T$ over the whole range. 
The qualitative trend is the same in both cases, with the strength of the quadrupole moment underestimated in the region of low strength,
and overestimated in the region of large strength, and the overall agreement is only slightly improved with a change in the slope. Similar to the
fixed charge models the scatter is substantial, and the RMS difference is found to be $\approx 10$\%, on average.

While the simple M-site models capture the overall qualitative trend in the change of the principal 
quadrupole moment over a broad range of configurations, an analysis of the RMS difference of the 
quadrupole moment components
shows that they deviate significantly for monomer configurations different than the ground state 
configuration. Neither the fixed charge or DMS charge M-site models (scaled or not) seem to better capture the principal quadrupole component over the other, and in all cases the RMS difference is around 10\% or greater. This illustrates that a three site model based on the M-site principle is not able to capture the variation of the quadrupole moment in a flexible water potential model to a good degree.
The four site QMS model developed in this work, which captures the principal quadrupole moment with a mean absolute error of 0.04 D\AA, similarly has low scatter throughout the range with an average RMS difference of 1.6$\%$, with greatest discrepancy in the region where the quadrupole moment interaction is the weakest.

Furthermore, the intermolecular interactions of the SCME/f model only depend on
five parameters. 
The parameters have been fitted to reproduce high level quantum chemistry calculations for the water dimer  
energy surface near the equilibrium geometry and interaction energy of the lowest-lying 
water clusters up to and including the hexamer, 
as well as the properties of the Ih ice crystal -- and in such a way that experimental values are reproduced to a good degree after including zero point energy corrections.

The simple parameterization of the flexible model and the use of a single center for 
the electrostatic 
interactions allows for the seamless integration into 
our recently implemented PE-QM/MM interface\cite{peqmmm1,peqmmm2}.

The calculated energy of the higher lying energy isomers of hexamer water cluster are found to be in a reasonable agreement with the results of CCSD(T) calculations in the complete basis set limit.\cite{bates:2009} The relative trend in the energy differences between the isomers, as well as the 
overall structures are captured to a good degree. 
This represents a significant improvement over the rigid SCME potential function and is on par with the trend predicted with the HBB2-pol \cite{hbbpola,hbbpolb} potential function, which explicitly models the N-body expansion up to the three-body terms in the interaction energy and is the predecessor of the MB-pol potential function.\cite{mbbpola,mbbpolb,mbbpolc} However, discrepancies are present in the series
of pentamer isomers, in particular the cage-like isomers. H-bonds in bonds where the distance
is greater than 3 \AA\ are found to be unstable, leading to a rearrangement of some of the SCME/f
structures compared to the reference structures.

Analysis of the vibrational modes of the small water clusters reveal a substantial 
improvement with the QMS mapping included (as opposed to a fixed value). In particular
are the intramolecular bending modes in the range 1600-1800 cm$^{-1}$, with 
maximum absolute deviation consistently less than 20 cm$^{-1}$ with the QMS included, 
compared to near-CBS CCSD(T) calculations.\cite{howardvib1,howardvib2} 
Importantly, including the DMS only results in the opposite trend of the intramolecular angle widening in crystal ice Ih compared to the gas phase. The inclusion of the QMS recovers the correct trend due to the balance between the magnitude of the dipole and principal quadrupole moment which are functions of the internal geometry and strongly dependent on this angle.

While the results presented here represent an important step forward in the development of a single center multipole expansion model for water, 
there is room for improvement, and this will be addressed in
future work. A natural next step to the mapping of the dipole and the quadrupole is to incorporate a mapping of the polarizability tensors. Work is ongoing
to incorporate the intramolecular geometry dependent mapping of the dipole-dipole, dipole-quadrupole and quadrupole-quadrupole polarizability tensors by Loboda et. al. \cite{loboda2016}.
It has been suggested that a critical part of the H-bond OH softening lies in the correct mapping of the polarizability surface of the individual monomers.\cite{burnham2008vibrational}

In particular, and in order to further address the overestimated H-bonded OH stretches, 
an improvement of the underlying water monomer
potential energy surface -- whose limit in terms of hydrogen dissociation is {\ce{OH^{.}} + \ce{H^{.}}} -- must be made when there are neighboring 
water molecules such that it approaches to some degree the dissociation limit in a condensed 
phase which is
{\ce{OH^{-}} + \ce{H^{+}}}. In order to capture this one could modify
the DMS and QMS charges to better represent this limit, and in a way which depends
on the environment. Modifying the charge of the DMS has, for example, previously been considered 
in water potentials in order to capture the charge delocalization and resulting softening of the H-bond, such as in the TTM3-F model.\cite{fanourgakis2008development}

Further improvements to this flexible SCME model that are being pursued include a more elaborate repulsive part including deviations from spherical symmetry. 


\begin{acknowledgement}
This work was supported by the University of Iceland Research Fund and the Icelandic Research Fund, grants no. 174082-051, 141080-051 and 207283-051. 
MG acknowledges post-doctoral fellowship from the University of Iceland Research Fund and thanks Ragnar Bj{\"o}rnsson for helpful discussions and guidance in the electronic structure calculations of the H$_2$O molecule.
JM acknowledges support from The Netherlands Organization for Scientific Research (NWO) under Vidi Grant No. 723.014.009.
Figures showing water molecules were drawn with the open source software Inkscape\cite{Inkscape} (licence GPL). 
\end{acknowledgement}


\begin{suppinfo}
The supporting information includes a detailed derivation of the atomic forces 
corresponding to contributions presented in eq~\eqref{eq:force-terms}, 
as well as a comparison between the numerical and analytical forces 
as the convergence criteria of the 
induced moments is varied. The parameters used for the model M-site description of the principal quadrupole moment are presented, followed by an analysis 
of the RMSD between ab initio versus the QMS quadrupole as well 
as model M-site quadrupoles with respect to geometrical variation of the 
monomer. Binding energies and relative structural properties
of the lowest-lying water clusters are given for the case where 
the quadrupole moment is set to a fixed value corresponding to 
the ground state monomer configuration. Finally, the evaluation
of the bulk properties from fitting the energy-volume relation --
with and without zero-point energy corrections -- is described.
\end{suppinfo}


\newpage


\begin{table*}
\caption{Numerical values and units of the quadrupole moment surface function, eq~\eqref{eq:QMS}.}
\begin{tabular}{lS|lS} 
\hline
Geometry &  & Charges & \\  
\hline
 a [\AA] & 0.5149 			& A & 0.9763 \\ 
 b       & -1.1271 			& B & 0.6418 \\
 c [\AA$^{-1}$] & 0.5146 	& C & 0.7251 \\
 d [rad] & 3.5908 			& D & -1.0603 \\
 e & -0.1081 				& $q^\mr{H}_\mr{eq}$ & 0.3310 \\
 $r_\mr{eq}$ [\AA] & 0.9578 & & \\
 $\theta_\mr{eq}$ [rad] & 1.8240 & & \\
\hline
\end{tabular}
\label{tbl:qms-para}
\end{table*}



\begin{center}
\begin{table*}
\caption{Properties of crystal ice Ih evaluated with SCME\cite{scme,wikfeldt:2013} and SCME/f, compared to experimental values. $\braket{r_\mr{OO}}$ is the average oxygen-oxygen distance, $a$, $b$, $c$ the lattice parameters for a dipole-free orthorhombic cell (containing eight molecules). 
$V_0^\mr{ZPE}$ ($V_0$) is the optimized cell volume, $E_\mr{lat}^\mr{ZPE}$ ($E_\mr{lat}$) and $B_0^\mr{ZPE}$ ($B_0$) are the lattice energy and bulk modulus with (and without) zero-point energy correction, all expressed per molecule.}
\begin{threeparttable}
\begin{tabular*}{.7\textwidth}{l@{\extracolsep{\fill}}|SSS} 
\hline
\multicolumn{1}{l|}{Property} & 
\multicolumn{1}{c}{SCME} & 
\multicolumn{1}{c}{SCME/f} & 
\multicolumn{1}{c}{Exp.\tnote{1}} \\
\hline
$\braket{r_\mr{OO}}$ [\AA]	&  2.742  &  2.751  &  2.751 \\ 
$a$ [\AA]                   &  4.470  &  4.478  &  4.497 \\
$b$ [\AA]                   &  7.747  &  7.777  &  7.789 \\
$c$ [\AA]                   &  7.287  &  7.331  &  7.321 \\
$V_0$ [\AA$^3$]             & 31.55   & 30.38   & \\
$V_0^\mr{ZPE}$ [\AA$^3$]    &         & 31.98   & 32.05 \\
$E_\mr{lat}$ [eV]           & -0.611  & -0.645  & -0.611 \\
$E_\mr{lat}^\mr{ZPE}$ [eV]  &         & -0.489  & -0.491 \\
$B_0$ [GPa]                 & 11.4    & 15.0    & \\
$B_0^\mr{ZPE}$ [GPa]        &         & 12.2    & 10.9 \\
\hline
\end{tabular*}
\begin{tablenotes}
{\footnotesize \item{1} Experimental values: average oxygen-oxygen distance is from \citet{Bjerrum385}, lattice parameters from \citet{Rottger:sh0050} (and resulting $V_0^\mr{ZPE}$), enthalpy of vaporization ($E_\mr{lat}^\mr{ZPE}$) and lattice energy ($E_\mr{lat}$) from \citet{whalley:1957}, and bulk modulus from \citet{hobbs2010ice}.}  
\end{tablenotes}
\end{threeparttable}
\label{tbl:ice}
\end{table*}
\end{center}


\begin{table*}
\caption{Intermolecular interaction model parameters, numerical values and units.}
\begin{tabular}{lS|lS} 
\hline
Damping &  & Repulsion & \\  
\hline
$\tau_d$ [\AA$^{-1}$] 	& 7.5548 	& $A_\mr{rep}$ [eV] & 8149.63 \\ 
g [\AA]           		& 1.1045 	& k & 0.5515 \\
  						&  			& h [\AA$^{-1}$] & 3.4695 \\
\hline
\end{tabular}
\label{tbl:modelpara}
\end{table*}



\begin{table*}
\caption{Interaction energy (kcal/mol) and distances (\AA) between atoms in the 
most stable configuration of clusters (H$_2$O)$_n$ with n=2,$\dots$,6.
 $E_\mr{int}$ is the SCME/f calculated interaction energy of the clusters
 and $\Delta E_\mr{int}$ (kcal/mol) the difference 
 with respect to the CCSD(T) values.
$\braket{d\br_\mr{OO}}$, $\braket{d\br_\mr{OH}}$ and $\braket{d\br_\mr{O\cdots H}}$
 are the RMSD of the oxygen-oxygen neighbour distances, intramolecular oxygen-hydrogen bond lengths 
 of the donor-hydrogen and bonding oxygen$\cdots$hydrogen bond length distances, respectively, compared
 to the 
 CCSD(T) obtained
 structures.\cite{bates:2009,temelso:2011} 
 $\braket{d\br^a}$ is the overall RMSD of the relaxed SCME/f structure
 evaluated using the Kabsch algorithm\cite{kabsch:1976}. All bond related differences are in units \AA. $\braket{d\angle\mr{OHO}}$ is the RMSD of the angle (in degrees) between the oxygen-hydrogen-oxygen in hydrogen bonds.
 }
\begin{tabular*}{\textwidth}{l@{\extracolsep{\fill}}|SSSSSSS}
\hline
\multicolumn{1}{l|}{(H$_2$O)$_n$}& 
\multicolumn{1}{c}{$E_\mr{int}$} & 
\multicolumn{1}{c}{$\Delta E_\mr{int}$} &
\multicolumn{1}{c}{$\braket{d\br_\mathrm{OO}}$} & 
\multicolumn{1}{c}{$\braket{d\br_\mathrm{OH}}$} & 
\multicolumn{1}{c}{$\braket{d\br_\mathrm{O\cdots H}}$} & 
\multicolumn{1}{c}{$\braket{d\br^a}$} & 
\multicolumn{1}{c}{$\braket{d\angle\mr{OHO}}$} \\
\hline
  2-Cs        &  -4.85  & +0.18  & 0.011 & 0.000 & 0.017 & 0.017 & 5.923 \\ 
  3-UUD       & -15.16  & +0.54  & 0.035 & 0.010 & 0.037 & 0.037 & 2.489 \\
  4-S4        & -27.51  & -0.11  & 0.005 & 0.014 & 0.006 & 0.045 & 1.382 \\
  5-CYC       & -36.72  & -0.71  & 0.014 & 0.015 & 0.003 & 0.046 & 0.369 \\
  6-PRI       & -47.10  & -1.18  & 0.017 & 0.012 & 0.035 & 0.033 & 4.564 \\
\hline
\end{tabular*}
\label{tbl:clusters}
\end{table*}



\begin{table*}[!th]
\caption{Energies and relative enenergy and structural properties of the pentamer and hexamer isomers. See the caption of Table \ref{tbl:clusters} for the definition of the table entries.}
\begin{threeparttable}
\begin{tabular*}{\textwidth}{l@{\extracolsep{\fill}}|SSSSSSS}
\hline
\multicolumn{1}{l|}{(H$_2$O)$_n$}& 
\multicolumn{1}{c}{$E_\mr{int}$} & 
\multicolumn{1}{c}{$\Delta E_\mr{int}$} &
\multicolumn{1}{c}{$\braket{d\br_\mathrm{OO}}$} & 
\multicolumn{1}{c}{$\braket{d\br_\mathrm{OH}}$} & 
\multicolumn{1}{c}{$\braket{d\br_\mathrm{O\cdots H}}$} & 
\multicolumn{1}{c}{$\braket{d\br^a}$} & 
\multicolumn{1}{c}{$\braket{d\angle\mr{OHO}}$} \\
\hline
  5-FRB       & -35.60 & -0.72 & 0.026 & 0.012 & 0.041 & 0.036 & 4.008 \\
  5-CAC       & -35.50 & -0.81 & 0.053 & 0.012 & 0.089 & 0.136 & 8.548 \\
  5-CAA       & -35.07 & -0.53 & 0.060 & 0.012 & 0.053 & 0.254 & 9.281 \\
  5-CAB       & -35.06 & -1.23 & 0.065 & 0.011 & 0.080 & 0.237 & 6.107 \\
  5-FRC       & -33.56 & -1.12 & 0.025 & 0.013 & 0.026 & 0.043 & 1.859 \\
  5-FRA       & -32.91 &  0.22 & 0.025 & 0.013 & 0.032 & 0.059 & 1.731 \\
\hline
  6-CAG       & -46.44 & -0.74 & 0.013 & 0.017 & 0.019 & 0.054 & 1.607 \\
  6-BK1       & -46.37 & -1.09 & 0.014 & 0.015 & 0.009 & 0.033 & 2.346 \\
  6-BK2       & -46.26 & -1.35 & 0.014 & 0.016 & 0.008 & 0.038 & 3.433 \\
  6-BAG       & -45.90 & -1.52 & 0.015 & 0.017 & 0.012 & 0.065 & 3.826 \\
  6-CYR       & -45.36 & -1.00 & 0.012 & 0.015 & 0.006 & 0.018 & 3.913 \\
  6-CB1       & -44.57 & -1.23 & 0.013 & 0.015 & 0.003 & 0.031 & 2.924 \\
  6-CB2       & -44.49 & -1.20 & 0.013 & 0.015 & 0.005 & 0.025 & 2.284 \\
\hline
\end{tabular*}
\begin{tablenotes}
{\footnotesize \item{1} Pentamers; fused-ring-B (FRB), cage-C
(CAC), cage-A (CAA), cage-B (CAB), fused-ring-C (FAC) and fused-ring-A (FRA); and the 
hexamers; cage (CAG), book-1 (BK1), book-2 (BK2),
bag (BAG), cyclic-ring (CYR), cyclic-boat-1 (CB1) and cyclic-boat-2 (CB2).}
\end{tablenotes}
\end{threeparttable}
\label{tbl:pentamers-hexamers}
\end{table*}



\begin{table*}
\caption{Relative vibrational properties of the lowest lying water clusters, including the cyclic ring isomer of the water hexamer cluster. 
The entries for each system correspond from top to bottom, the low-frequency intermolecular vibrational modes (l, 10-1000 cm$^{-1}$), intramonomer bending (b, 1600-1800 cm$^{-1}$) and high frequency stretching of H-bond OH and non-bonded OH bonds (h, 3200-3900 cm$^{-1}$).
$\braket{\Delta\mr{cm}^{-1}}$ is the RMSD between the frequencies in the low, medium and high range, as predicted with SCME/f compared to near-CBS CCSD(T) reference calculations.\cite{howardvib1,howardvib2} The last entry is RMSD for the total frequency range ({\bf t}), where the value in the parenthesis excludes the overestimated H-bond OH stretches. 
max$\mid\! \Delta\mr{cm}^{-1}\!\mid$ is the maximum absolute difference for each entry.
The two columns on the right are for the SCME/f model potential with the quadrupole moment
set to a fixed value corresponding to the ground state water monomer configuration.}
\begin{tabular*}{\textwidth}{l@{\extracolsep{\fill}}c|SlS|SlS} 
\hline
& & 
\multicolumn{3}{c|}{quadrupole moment surface} &
\multicolumn{3}{c}{fixed quadrupole moment} 
\\
\multicolumn{1}{l}{(H$_2$O)$_n$} & 
\multicolumn{1}{c|}{} &
\multicolumn{1}{c}{$\braket{\Delta\mr{cm}^{-1}}$} & &
\multicolumn{1}{c|}{max$\mid\! \Delta\mr{cm}^{-1}\!\mid$} &
\multicolumn{1}{c}{$\braket{\Delta\mr{cm}^{-1}}$} & &
\multicolumn{1}{c}{max$\mid\! \Delta\mr{cm}^{-1}\!\mid$} 
\\
\hline
  2-Cs      & l  & 17.52 & & 41.50   & 15.33 & & 23.10 \\
            & b  &  6.71 & &  9.10   &  9.18 & & 12.70 \\
            & h  & 28.55 & & 36.10   & 50.41 & & 70.30 \\
            & {\bf t} & {\bf 17.23} & & 
                      & {\bf 23.56} & & \\
\hline
  3-UUD     & l  & 23.03 & & 49.20 &  34.43 & &  71.70 \\
            & b  &  7.75 & & 12.20 &  18.41 & &  23.90 \\
            & h  & 68.09 & & 89.30 & 153.65 & & 190.40 \\
            & {\bf t} & {\bf 31.20} & (24.98) & 
                      & {\bf 64.02} & (29.19) & \\
\hline
  4-S4      & l  &  20.87 & &  44.40  &  29.43 & &  58.10 \\
            & b  &  10.74 & &  12.80  &  25.69 & &  29.40 \\
            & h  & 155.58 & & 200.20  & 266.11 & & 334.10 \\
            & {\bf t} & {\bf  59.19} & (21.06) &    
                      & {\bf 100.25} & (26.49) & \\ 
\hline
  5-CYC     & l  &  18.47 & &  35.20  &  35.08 & &  65.30 \\
            & b  &  14.22 & &  19.60  &  26.44 & &  34.40 \\
            & h  & 179.34 & & 229.60  & 282.41 & & 358.10 \\
            & {\bf t} & {\bf  66.02} & (20.13) & 
                      & {\bf 105.22} & (31.38) & \\
\hline
  6-CYR     & l  &  21.10 & &  44.30  &  35.16 & &  75.40 \\
            & b  &  11.55 & &  13.10  &  32.53 & &  40.00 \\
            & h  & 185.38 & & 239.30  & 280.06 & & 356.30 \\
            & {\bf t} & {\bf  67.75} & (23.24) & 
                      & {\bf 103.48} & (32.56) & \\ 
\hline
  6-PRI     & l  &  21.70 & &  87.60  &  27.74 & &  56.20 \\
            & b  &   9.86 & &  18.30  &  35.16 & &  45.80 \\
            & h  & 208.58 & & 408.60  & 313.99 & & 571.80 \\
            & {\bf t} & {\bf  75.79} & (24.60) & 
                      & {\bf 113.85} & (36.40) & \\
\hline
\end{tabular*}
\label{tbl:vibrations}
\end{table*}


\begin{table*}
\caption{Average intramolecular HOH angles (in degrees) for the SCME/f model with and without the QMS. Experimental angles for the isolate water molecule (gas) and in crystal ice Ih (Ih) are presented for comparison.
 }
\begin{tabular*}{\textwidth}{l@{\extracolsep{\fill}}|SSSS}
\hline
\multicolumn{1}{l|}{}& 
\multicolumn{1}{c}{Exp (gas)} & 
\multicolumn{1}{c}{Exp (Ih)} &
\multicolumn{1}{c}{SCME/f } & 
\multicolumn{1}{c}{SCME/f no QMS} \\
\hline
 $\braket{\angle \mr{HOH}}$  &  104.5  &  108.1  &  106.51  &  99.95   \\ 

\hline
\end{tabular*}
\label{tbl:angles}
\end{table*}
 

\clearpage

\bibliography{ms}

\begin{mcitethebibliography}{118}
\bibitem[Lin and Truhlar(2006)Lin, and Truhlar]{lin:2006}
Lin,~H.; Truhlar,~D.~G. QM/MM: what have we learned, where are we, and where do
  we go from here? \emph{Theor. Chem. Acc.} \textbf{2006}, \emph{117},
  185\relax
\mciteBstWouldAddEndPuncttrue
\mciteSetBstMidEndSepPunct{\mcitedefaultmidpunct}
{\mcitedefaultendpunct}{\mcitedefaultseppunct}\relax
\EndOfBibitem
\end{mcitethebibliography}

\end{document}


\tableofcontents
\clearpage

\section{Analytical Forces}

We further apply the chain rule considering the main electrostatic 
force expression in the main text
\begin{align}
    -\left(\frac{\partial E_\mr{in+pol}}{\partial r^{ia}_\al} + \frac{\partial E_\mr{self}}{\partial r^{ia}_\al}\right) =&
    -\frac{\partial E_\mr{in+pol}}{\partial \mu^j_\be(\{\br^{jb}\})}
     \frac{\partial \mu^j_\be(\{\br^{jb}\})}{\partial r^{ia}_\al} 
    -\frac{\partial E_\mr{in+pol}}{\partial \theta^j_\by(\{\br^{jb}\})}
     \frac{\partial \theta^j_\by(\{\br^{jb}\})}{\partial r^{ia}_\al}
     \nonumber \\
    &-\frac{\partial E_\mr{in+pol}}{\partial V^j_{\byde\dots\eta}}
     \frac{\partial V^j_{\byde\dots\eta}}{\partial r^{ia}_\al}
     -\left(\frac{\partial E_\mr{in+pol}}{\partial R^j_{\nu o}} + 
     \frac{\partial E_\mr{self}}{\partial R^j_{\nu o}}
      \right)\frac{\partial R^j_{\nu o}}{\partial r^{ia}_\al} \label{eq:f-tot}
\end{align}
The last term on the right hand side describes the force contribution due to
the definition of the local-to-global reference frame transformation, and 
is the only term which includes an explicit contribution to the atomic forces
due to the self-energies. To see this we first write the MM induced dipoles and quadrupoles as
\begin{align}
\Delta\mi_\al = -\alpha_\ab V^i_\be 
                - \frac{1}{3}A_{\aby} V^i_\by = \Delta\mi_\al(\al) + \Delta\mi_\al(A)\\
\Delta\qi_\ab = -A_{\gamma\ab} V^i_\gamma 
                   - C_{\gamma\delta\ab} V^i_{\gamma\delta} = \Delta\qi_\ab(A) + \Delta\qi_\ab(C)
\end{align}
where on the right hand side of the second equality the contribution from the external field and field gradient due to the on-site potential is split up. 
With these definitions it is easy to relate the external field and field gradient at site $i$ to the self-consistent moments of 
molecule $i$
\begin{align}\label{eq:indtopot}
    V^i_\be =& -\frac{\Delta\mi_\al(\al)}{\al_\ab} \\
    V^i_\gamma =& -\frac{\Delta\qi_\ab(A)}{A_{\gamma\ab}} \label{eq:indtopotquad}
\end{align}
and
\begin{align}\label{eq:indtopotgrad}
    V^i_{\gamma\delta} =& -\frac{\Delta\qi_\ab(C)}{C_{\gamma\delta\ab}} \\
    V^i_\by =& -\frac{\Delta\mi_\al(A)}{A_{\aby}} \label{eq:indtopotgraddip} .
\end{align}
The self-energy of an induced dipole in linear response theory is 
\begin{equation}
    E^\mu_\mr{self} = -\int_{0}^{\Delta\mi} V^i_\be d\Delta\mi .
\end{equation}
It gives the energy cost of inducing a first order moment in the potential field at site $i$. By inserting the relation in eq~\eqref{eq:indtopot} into the equation above, and by considering only (for the moment) the induced dipole in in response to an external field gives
\begin{equation}
    E^\mu_\mr{self} = \int_0^{\Delta\mi(\al)}\frac{\Delta\mi_\al(\al)}{\al_\ab}d\Delta\mi 
    = \frac{1}{2}\frac{\Delta\mi_\al(\al)\Delta\mi_\be(\al)}{\al_\ab} .
\end{equation}
For isotropic atomic polarization this becomes
\begin{equation}
    E_\mr{self}^\mr{iso} = \frac{1}{2}\frac{(\Delta\mi)^2}{\al} .
\end{equation}
This form is frequently encountered in MM work based on isotropic atomic polarization and induced dipole in response to an external field.
Similarly for the induced quadrupole 
\begin{equation}
    E^\theta_\mr{self} = -\frac{1}{3}\int_0^{\Delta\qi}V^i_\by d\Delta\qi
\end{equation}
which expresses the energy cost of inducing a second order moment in the field gradient at site $i$. The factor of $1/3$ follows from the definition of the traceless Cartesian moments\cite{stone2013theory} used in SCME. The total self-energy for a single site $i$ in SCME is then
\begin{align}
    E_\mr{self} =& E^\mu_\mr{self} + E^\theta_\mr{self} \nonumber \\
    =& \int_0^{\Delta\mi}\frac{\Delta\mi_\al(\al)}{\al_\ab}d\Delta\mi 
    + \frac{1}{3}\int_0^{\Delta\qi}\frac{\Delta\qi_\ab(C)}{C_{\gamma\delta\ab}}d\Delta\qi \nonumber \\
    =& \frac{1}{2}\frac{\Delta\mi_\al(\al)}{\al_\ab}\left(\Delta\mi_\be(\al) 
    +\frac{1}{3}\Delta\mi_\be(A)\right) + \frac{1}{6}\frac{\Delta\qi_\ab(C)}{C_{\gamma\delta\ab}}\left(\Delta\qi_{\gamma\delta}(C)
    +\Delta\qi_{\gamma\delta}(A)\right) \nonumber \\
    =& \frac{1}{2}\frac{\Delta\mi_\al(\al)\Delta\mi_\be(\al)}{\al_\ab} 
     + \frac{1}{3}\frac{\Delta\mi_\al(\al)\Delta\qi_\by(C)}{k_{\aby}}
     + \frac{1}{6}\frac{\Delta\qi_\ab(C)\Delta\qi_{\gamma\delta}(C)}{C_{\gamma\delta\ab}}
     \label{eq:selfenergy}
\end{align}
where the relations in eqs~\eqref{eq:indtopot}--\eqref{eq:indtopotgraddip} are used. The matrix $k$ is given by
\begin{equation}
 k = \frac{\al C}{A}
\end{equation}
This expression for the self-energies is very useful at self-consistency (SCF). First and foremost it shows that there are no force contributions arising from partial derivatives of the on-site potential field and field gradients when considering the self-energy terms, since
at SCF we have
\begin{equation}
    \frac{\partial E_\mr{tot}}{\partial \Delta\mi_\al} =  \frac{\partial E_\mr{tot}}{\partial \Delta\qi_\al} = 0,
\end{equation}
which implies 
\begin{equation}
 \frac{\partial E_\mr{self}}{\partial V^{jb}_{\byde\dots\eta}} = 0
\end{equation}

Contributions arise from the static
octupole and static hexadecapole, as well as the dipole-dipole, dipole-quadrupole
quadrupole-quadrupole polarizability matrices. The general
contributions are of the following form
\begin{equation}
    \frac{\partial E}{\partial R^j_{\nu o}}
     \frac{\partial R^j_{\nu o}}{\partial r^{ia}_\al} \to \delta_{ij}\delta_{\nu o,\beta\gamma} 
\end{equation}
resulting in for the fixed moments
\begin{align}
    \frac{\partial E_\mr{in+pol}}{\partial R^j_{\nu o}}\frac{\partial R^j_{\nu o}}{\partial r^{ia}_\al} =& \frac{1}{15}\left(\frac{\partial R^i_{\eta\be}}{\partial r^i_\al}R^i_{\tau\gamma}R^i_{\kappa\delta} + 
    R^i_{\eta\be}\frac{\partial R^i_{\tau\gamma}}{\partial r^i_\al}R^i_{\kappa\delta}
    + R^i_{\eta\be}R^i_{\tau\gamma}
    \frac{\partial R^i_{\kappa\delta}}{\partial r^i_\al}
    \right)\oil_{\eta\tau\kappa}V^i_{\be\gamma\delta} \nonumber \\
    &+ \frac{1}{105}\Bigg(
    \frac{\partial R^i_{\eta\be}}{\partial r^{ia}_\al}R^i_{\tau\gamma} R^i_{\kappa\delta}R^i_{\sigma\eta} + 
    R^i_{\eta\be}\frac{\partial R^i_{\tau\gamma}}{\partial r^{ia}_\al} R^i_{\kappa\delta}R^i_{\sigma\eta} \nonumber \\
    &\ \ \ \ \ \ + R^i_{\eta\be}R^i_{\tau\gamma}
    \frac{\partial R^i_{\kappa\delta}}{\partial r^{ia}_\al}R^i_{\sigma\eta}
    + R^i_{\eta\be}R^i_{\tau\gamma}
    R^i_{\kappa\delta}\frac{\partial R^i_{\sigma\eta}}{\partial r^{ia}_\al}
    \Bigg)\hil_{\eta\tau\kappa\sigma}V^i_\byde
\end{align}
and for the polarizability matrices the contributions are
\begin{align}
   &\left(\frac{\partial E_\mr{in+pol}}{\partial R^j_{\nu o}} + 
   \frac{\partial E_\mr{self}}{\partial R^j_{\nu o}}\right)\frac{\partial R^j_{\nu o}}{\partial r^{ia}_\al} = -\frac{1}{2}\left(\frac{\partial 
    R^i_{\eta\be}}{\partial r^i_\al}R^i_{\tau\gamma}
    + R^i_{\eta\be}\frac{\partial R^i_{\tau\gamma}}{\partial r^i_\al}
    \right)\ddl_{\eta\tau}V^i_\be V^i_\gamma \nonumber \\
    &-\frac{1}{3}\left(\frac{\partial R^i_{\eta\be}}{\partial r^i_\al}R^i_{\tau\gamma}R^i_{\kappa\delta} + 
    R^i_{\eta\be}\frac{\partial R^i_{\tau\gamma}}{\partial r^i_\al}R^i_{\kappa\delta}
    + R^i_{\eta\be}R^i_{\tau\gamma}
    \frac{\partial R^i_{\kappa\delta}}{\partial r^i_\al}\right)
    \dql_{\eta\tau\kappa}V^i_\be V^i_{\gamma\delta} \nonumber \\
    &-\frac{1}{6}\Bigg(
    \frac{\partial R^i_{\eta\be}}{\partial r^i_\al}R^i_{\tau\gamma}
    R^i_{\kappa\delta}R^i_{\sigma\eta} + 
    R^i_{\eta\be}\frac{\partial R^i_{\tau\gamma}}{\partial r^i_\al}
    R^i_{\kappa\delta}R^i_{\sigma\eta} \nonumber \\
    &\ \ \ \ \ + R^i_{\eta\be}R^i_{\tau\gamma}
    \frac{\partial R^i_{\kappa\delta}}{\partial r^i_\al}R^i_{\sigma\eta}
    +R^i_{\eta\be}R^i_{\tau\gamma}
    R^i_{\kappa\delta}\frac{\partial R^i_{\sigma\eta}}{\partial r^i_\al}
    \Bigg)\qql_{\eta\tau\kappa\sigma}V^i_\by V^i_{\delta\eta}
\end{align}
where the factors $1/2$, $1/3$ and $1/6$ are due to the self-energy terms -- i.e. the 
contribution from the electrostatic interaction is reduced by exactly one-half
due to net cancellation by the self-energy terms.

Different
choices of local frames and principal vectors, as well as atomic force contributions,
are detailed in the work of Lipparini et. al.\cite{lipparini2014scal} The specific choices 
in this work which define the rotation matrix $R^i_{\Lambda\lambda}$ result in obvious sign changes compared to their work, so the atomic contributions are detailed below in compact form. We require the
derivatives of the rotation matrix with respect to the atomic coordinates which can be written as
\begin{equation}
    \frac{\partial R^i_{\Lambda\lambda}}{\partial r^{ia}_\al} = \begin{bmatrix}
             \frac{\partial e^{iX}_x}{\partial r^{ia}_\al} & 
             \frac{\partial e^{iX}_y}{\partial r^{ia}_\al} & 
             \frac{\partial e^{iX}_z}{\partial r^{ia}_\al} \\
             \frac{\partial e^{iY}_x}{\partial r^{ia}_\al} & 
             \frac{\partial e^{iY}_y}{\partial r^{ia}_\al} & 
             \frac{\partial e^{iY}_z}{\partial r^{ia}_\al} \\
             \frac{\partial e^{iZ}_x}{\partial r^{ia}_\al} & 
             \frac{\partial e^{iZ}_y}{\partial r^{ia}_\al} & 
             \frac{\partial e^{iZ}_z}{\partial r^{ia}_\al} \\
            \end{bmatrix} \label{eq:derrot}
\end{equation}
The \ecom\ and principal vectors used to define the rotation are
\begin{equation}
    \br^i = \sum_{a\in i}\br^{ia}\frac{M^a}{M^i},\ \ \ \ 
    {\bf B}^i = \br^i - \br^{i{\mr{H}_1}},\ \ \ \
    {\bf C}^i = \br^i - \br^{i{\mr{H}_2}}
\end{equation}
Defining
\begin{equation}
    {\bf D}^i = B^i{\bf C^i} + C^i{\bf B^i}
\end{equation}
such that 
\begin{equation}
    {\bf e}^{iZ} = \frac{{\bf D}^i}{D^i}
\end{equation}
where $B^i, C^i$ and $D^i$ are the euclidean norms of the vectors, the terms in the derivative of the rotation matrix, eq~\eqref{eq:derrot}, are then
\begin{align}
    \frac{\partial e^{iZ}_\lambda}{\partial r^{ia}_\al} =& 
    \frac{\partial e^{iZ}_\lambda}{\partial D^i_\be}\left(
    \frac{\partial D^i_\be}{\partial r^i_\gamma}
    \frac{\partial r^i_\gamma}{\partial r^{ia}_\al}+\frac{\partial D^i_\be}{\partial r^{ia}_\al}\right)
    \\
    \frac{\partial e^{iX}_\lambda}{\partial r^{ia}_\al} =& 
    \frac{\partial e^{iX}_\lambda}{\partial B^i_\be}\left(
    \frac{\partial B^i_\be}{\partial r^i_\gamma}
    \frac{\partial r^i_\gamma}{\partial r^{ia}_\al}
    + \frac{\partial B^i_\be}{\partial r^{ia}_\al}
    \right)
    +
    \frac{\partial e^{iX}_\lambda}{\partial e^{iZ}_\be}\frac{\partial e^{iZ}_\be}
    {\partial r^{ia}_\al} \\
    \frac{\partial e^{iY}_\lambda}{\partial r^{ia}_\al} =&
    \frac{\partial e^{iY}_\lambda}{\partial e^{iX}_\be}\frac{\partial e^{iX}_\be}
    {\partial r^{ia}_\al}
    +
    \frac{\partial e^{iY}_\lambda}{\partial e^{iZ}_\be}\frac{\partial e^{iZ}_\be}
    {\partial r^{ia}_\al}
\end{align}
where the leading terms are as follows
\begin{align}
    \frac{\partial e^{iZ}_\lambda}{\partial D^i_\be} =& \left(\frac{\bf I}{D^i} - \frac{{\bf D}^i\otimes{\bf D}^i}{(D^i)^3}\right)_{\lambda\be} \\
    \frac{\partial e^{iX}_\lambda}{\partial B^i_\be} =& \left(\frac{{\bf I} 
    - {\bf e}^{iZ}\otimes{\bf e}^{iZ} - {\bf e}^{iX}\otimes{\bf e}^{iX}}
    {\mid {\bf B}^i - ({\bf B}^i\cdot{\bf e}^{iZ}){\bf B}^i\mid}\right)_{\lambda\be} \\
    \frac{\partial e^{iX}_\lambda}{\partial e^{iZ}_\be} =& \left(\frac{
    ({\bf B}^i\cdot{\bf e}^{iZ}){\bf e}^{iX}\otimes{\bf B}^i}
    {\mid {\bf B}^i - ({\bf B}^i\cdot{\bf e}^{iZ}){\bf B}^i\mid^2}
    - \frac{({\bf B}^i\cdot{\bf e}^{iZ}){\bf I} + {\bf e}^{iZ}\otimes{\bf B}^i}
    {\mid {\bf B}^i - ({\bf B}^i\cdot{\bf e}^{iZ}){\bf B}^i\mid}\right)_{\lambda\be} \\
    \frac{\partial e^{iY}_\lambda}{\partial e^{iZ}_\be} =& 
    \epsilon_{\lambda\sigma\tau}\delta_{\be\sigma}e^{iX}_\tau \\
    \frac{\partial e^{iY}_\lambda}{\partial e^{iX}_\be} =& 
    \epsilon_{\lambda\sigma\tau}e^{iZ}_\sigma\delta_{\be\tau} 
\end{align}
where ${\bf I}$ is the $3\times3$ identity matrix and $\epsilon_\aby$ the Levi-Civita symbols. 
The latter terms are
\begin{align}
     \frac{\partial D^i_\be}{\partial r^{i}_\gamma}
     \frac{\partial r^i_\gamma}{\partial r^{ia}_\al} =& 
     \left((B^i + C^i){\bf I} + \frac{{\bf B}^i\otimes{\bf C}^i}{C^i}
     + \frac{{\bf C}^i\otimes{\bf B}^i}{B^i}\right)_{\be\gamma}
     \delta_{\gamma\al}\frac{M^a}{M^i} \\
     \frac{\partial D^i_\be}{\partial r^{i\mr{H}_1}_\al} =& 
     -\left(B^i{\bf I} + \frac{{\bf B}^i\otimes{\bf C}^i}{C^i}\right)_{\be\al} \\
     \frac{\partial D^i_\be}{\partial r^{i\mr{H}_2}_\al} =& 
     -\left(C^i{\bf I} + \frac{{\bf C}^i\otimes{\bf B}^i}{B^i}\right)_{\be\al} \\
     \frac{\partial B^i_\be}{\partial r^i_\gamma}
     \frac{\partial r^i_\gamma}{\partial r^{ia}} =& \delta_{\be\gamma}\delta_{\gamma\al}\frac{M^a}{M^i} \\
     \frac{\partial B^i_\be}{\partial r^{i\mr{H}_1}_\al} =& -\delta_{\be\al}
\end{align}
For the DMS (see the main text) the first term 
on the right hand side is
\begin{align}
    \frac{\partial E_\mr{in+pol}}{\partial \mu^j_\be(\{\br^{jb}\})} =& 
    \left(\frac{\partial E_\mr{in+pol}}{\partial \mu^j_\be(\{\br^{jb}\})} +
    \frac{\partial E_\mr{in+pol}}{\partial V^k_{\gamma\delta\dots\eta}}
    \frac{\partial V^k_{\gamma\delta\dots\eta}}{\partial \mu^j_\be(\{\br^{jb}\})}\right)
    = \frac{1}{2}V^j_\be + \frac{1}{2}\sum_k^n \delta_{jk} V^k_\be \\
    \frac{\partial \mu^j_\be(\{\br^{jb}\})}{\partial r^{ia}_\al} =&
    \left(\frac{\partial q^{jb}}{\partial r^{ia}_\al} 
    + \frac{\partial r^{jb}}{\partial r^{ia}_\al}\right) = 
    \delta_{ji}\left(\sum_{b}^{n_i}\frac{\partial q^{jb}}{\partial r^{ia}_\al}r^{jb}_\be + \delta_{ba}q^{jb}\delta_{\be\al}\right) \\
    \frac{\partial E_\mr{in+pol}}{\partial \mu^j_\be(\{\br^{jb}\})}
    \frac{\partial \mu^j_\be(\{\br^{jb}\})}{\partial r^{ia}_\al} =& q^{ia}V^i_\al 
    + \sum_{b}^{n_i}\frac{\partial q^{ib}}{\partial r^{ia}_\al}r^{ib}_\be V^i_\be
\end{align}
The derivatives of the DMS, $\frac{\partial q^{ib}}{\partial r^{ia}_\be}$, with respect to the atomic positions are derived by Burnham et al \cite{burnham2002development} and 
are available in open source repositories.

For the force contribution due to the QMS we first rewrite the following expression  
\begin{equation}
    \qi_\ab(\br^{i\mr{O}},\br^{i\mr{H}_1},\br^{i\mr{H}_2}) = \sum_{a}
    ^{\mr{H}_1',\mr{H}_2',\mr{L}_1,\mr{L}_2} 
    \frac{3}{2}\bigg\{q^{ia}\left(
    (\br^{ia}-\br^i)_\al(\br^{ia}-\br^i)_\be
    -\frac{\delta_\ab}{3}||\br^{ia}-\br^i||\right)\bigg\}
\end{equation}
noting that the position of the L-sites in the global frame are
\begin{equation}
    r^{i\mr{L}_l}_\al = R^{i\mr{H}_l}_{\eta\al}e^{iZ}_\eta f(\br^{\mr{H}_l}) + r^{i}_\al
\end{equation}
we remove redundant terms and the expression for the QMS becomes
\begin{align}
    \qi_\ab(\br^{i\mr{O}},\br^{i\mr{H}_1},\br^{i\mr{H}_2}) = \frac{3}{2}&\bigg\{\sum_{a}^{\mr{H}_1',\mr{H}_2'}q^{ia}\left(
    (\br^{ia}-\br^i)_\al(\br^{ia}-\br^i)_\be
    -\frac{\delta_\ab}{3}||\br^{ia}-\br^i||\right) \nonumber \\
    &+ \sum_{l}^{1,2} q^{i\mr{L}_l}\left(
    dr^{i\mr{L}_l}_\al dr^{i\mr{L}_l}_\be
    -\frac{\delta_\ab}{3}||dr^{i\mr{L}_l}||\right) 
    \bigg\}
\end{align}
where
\begin{equation}
    dr^{i\mr{L}_l}_\al = R^{i\mr{H}_l}_{\eta\al}e^{iZ}_\eta f(\br^{\mr{H}_l}) 
\end{equation}
Similar to the DMS we have
\begin{align}
    \frac{\partial E_\mr{in+pol}}{\partial \theta^j_\by(\{\br^{jb}\})} =& 
    \left(\frac{\partial E_\mr{in+pol}}{\partial \theta^j_\by(\{\br^{jb}\})} 
    + \frac{\partial E_\mr{in+pol}}{\partial V^k_{\delta\kappa\dots\eta}}
      \frac{\partial V^k_{\delta\kappa\dots\eta}}
      {\partial \theta^{j}_\by(\{\br^{jb}\})}\right) 
      = \frac{1}{6}V^j_\by + \frac{1}{6}\sum_k^n\delta_{jk}V^k_\by \\
    \frac{\partial \theta^j_\by(\{\br^{jb}\})}{\partial r^{ia}_\al} =&
    \frac{3}{2}\delta_{ji}\Bigg\{\sum_{b}^{\mr{H}_1',\mr{H}_2'}
    \frac{\partial q^{jb}}{\partial r^{ia}_\al}
    \left(
    (\br^{jb}-\br^j)_\be(\br^{jb}-\br^j)_\gamma
    -\frac{\delta_\by}{3}||\br^{jb}-\br^j||\right) \nonumber \\
    &\ \ \ \ + \sum_{l}^{1,2}\frac{\partial q^{j\mr{L}_l}}{\partial r^{ia}_\al}
    \left(
    dr^{j\mr{L}_l}_\be dr^{j\mr{L}_l}_\gamma
    -\frac{\delta_\by}{3}||dr^{j\mr{L}_l}||\right) \nonumber \\
    &\ \ \ \ + \sum_{b}^{\mr{H}_1',\mr{H}_2'}q^{jb}\Bigg(\delta_{\be\al}\left(\delta_{ba} - \sum_{c}^{n_j}\delta_{ca}\frac{M^c}{M^j}\right)(\br^{jb}-\br^j)_\gamma \nonumber
    \\
    &\ \ \ \ \ \ \ \ \ \ \ \ \ \ \ \ + (\br^{jb}-\br^j)_\be
    \delta_{\gamma\al}\left(\delta_{ba} - \sum_{c}^{n_j}\delta_{ca}\frac{M^c}{M^j}\right)
    \nonumber \\
    &\ \ \ \ \ \ \ \ \ \ \ \ \ \ \ \ +
    \delta_{\by}\frac{2}{3}\left(\delta_{ba} - \sum_{c}^{n_j}\delta_{ca}\frac{M^c}{M^j}\right)
    \delta_{\al\delta}(\br^{jb} - \br^j)_\delta
    \Bigg) \nonumber \\
    &\ \ \ \ + \sum_{l}^{1,2}q^{j\mr{L}_l}\Bigg(
    \delta_{\be\al}\frac{\partial dr^{j\mr{L}_l}_\be}{\partial r^{ia}_\al}dr^{j\mr{L}_l}_\gamma
    +dr^{j\mr{L}_l}_\be\delta_{\gamma\al}\frac{\partial dr^{j\mr{L}_l}_\gamma}
    {\partial r^{ia}_\al} \nonumber \\
    &\ \ \ \ \ \ \ \ \ \ \ \ \ \ \ \ +
    \delta_{\be\gamma}\frac{2}{3}\frac{\partial r^{j\mr{L}_l}_\delta}{\partial r^{ia}_\al}
    dr^{j\mr{L}_l}_\delta\Bigg)
    \Bigg\}
\end{align}

The terms involving the partial derivative of the charges for each site are 
readily available since through the
definition of the QMS charges we have
\begin{align}
    q^{i\mr{H}_l'} =& \mr{A}q^{i\mr{H}_l} + \mr{B}q^{i\mr{H}_l}_\mr{eq} \\
    q^{i\mr{L}_l}  =& \mr{C}q^{i\mr{H}_l} + \mr{D}q^{i\mr{H}_l}_\mr{eq}
\end{align}
and hence the expression reduces to
\begin{align}
    \frac{\partial E_\mr{in+pol}}{\partial \theta^j_\by(\{\br^{jb}\})}
    \frac{\partial \theta^j_\by(\{\br^{jb}\})}{\partial r^{ia}_\al} =&
    \frac{1}{2}\sum_l^{1,2}\Bigg\{ \frac{\partial q^{i\mr{H}_l}}{\partial r^{ia}_\al} 
    \Bigg(\mr{A}\bigg(dr^{i\mr{H}_l}_\be dr^{i\mr{H_l}}_\gamma 
    - \frac{\delta_{\by}}{3}||dr^{i\mr{H}_l}||\bigg)  \nonumber \\
    &+ \mr{C}\bigg(dr^{i\mr{L}_l}_\be dr^{i\mr{L}_l}_\gamma 
    - \frac{\delta_\by}{3}||dr^{i\mr{L}_l}||\bigg)
    \Bigg) \nonumber \\
    &+q^{i\mr{H}'_l}\Bigg(2\left(\delta_{ba} - \sum_c^{n_i}\delta_{ca}\frac{M^c}{M^i}\right)
    dr^{i\mr{H_l}}_\gamma\delta_{\be\al} \nonumber \\
    &\ \ \ \ \ \ \ - \frac{2}{3}\left(\delta_{ba} - \sum_c^{n_i}\delta_{ca}\frac{M^c}{M^i}\right)
    dr^{i\mr{H}_l}_\al \delta_\by \Bigg) \nonumber \\
    &+ q^{i\mr{L}_l}\Bigg(
    2\frac{\partial dr^{i\mr{L}_l}_\be}{\partial r^{ia}_\al}dr^{i\mr{L}_l}_\gamma 
    - \frac{2}{3}\frac{\partial dr^{i\mr{L}}_\delta}{\partial r^{ia}_\al}dr^{i\mr{L}_l}_\delta
    \delta_\by
    \Bigg)
    \Bigg\}V^{i}_\by
\end{align}
where 
\begin{equation}
    dr^{i\mr{H}_l}_\al = (\br^{i\mr{H}_l} - \br^i)_\al
\end{equation}

The only unknowns are the derivatives of the position of the L-sites in 
the local frame reference. Applying the rotation operators
\begin{align}
    {\bf R}^{i\mr{L}_1} =& \left( \mr{cos}(\theta'){\bf I} -
    \mr{sin}(\theta')\left[{\bf e}^{iX}\right]_\times\right) \\
    {\bf R}^{i\mr{L}_2} =& \left( \mr{cos}(\theta'){\bf I} +
    \mr{sin}(\theta')\left[{\bf e}^{iX}\right]_\times\right)
\end{align}
on ${\bf e}^{iZ}$ in the case of L$_1$ and L$_2$ 
the expression for the local frame vectors becomes
\begin{align}
    dr^{i\mr{L}_1}_\al = \left(\mr{cos}(\theta')e^{iZ}_\al
    - \mr{sin}(\theta')e^{iY}_\al\right)f(\br^{i\mr{H}_1}) \\
    dr^{i\mr{L}_2}_\al = \left(\mr{cos}(\theta')e^{iZ}_\al
    + \mr{sin}(\theta')e^{iY}_\al\right)f(\br^{i\mr{H}_2})
\end{align}
Derivatives of the expressions above are of the form
\begin{align}
    \frac{\partial dr^{i\mr{L}_l}_\be}{\partial r^{ia}_\al} =&
    \frac{\partial dr^{i\mr{L}_l}_\be}{\partial e^{iZ}_\gamma}
    \frac{\partial e^{iZ}_\gamma}{\partial r^{ia}_\al}
    +
    \frac{\partial dr^{i\mr{L}_l}_\be}{\partial e^{iY}_\gamma}
    \frac{\partial e^{iY}_\gamma}{\partial r^{ia}_\al}
    +
    \frac{\partial dr^{i\mr{L}_l}_\be}{\partial f(\br^{\mr{H}_l})}
    \frac{\partial f(\br^{\mr{H}_l})}{\partial r^{ia}_\al} \nonumber \\
    &+ \left(\frac{\partial dr^{i\mr{L}_l}_\be}{\partial \mr{cos}(\theta')}
       \frac{\partial \mr{cos}(\theta')}{\partial \theta'}
     +\frac{\partial dr^{i\mr{L}_l}_\be}{\partial \mr{sin}(\theta')}
       \frac{\partial \mr{sin}(\theta')}{\partial \theta'}
       \right)\frac{\partial \theta'}{\partial r^{ia}_\al}
\end{align}
For $l=1$ as an example the leading terms are
\begin{align}
    \frac{\partial dr^{i\mr{L}_l}_\be}{\partial e^{iZ}_\gamma} =& \mr{cos}(\theta')
    f(\br^{i\mr{H}_1})\delta_{\be\gamma} \\
    \frac{\partial dr^{i\mr{L}_l}_\be}{\partial e^{iY}_\gamma} =& -\mr{sin}(\theta')
    f(\br^{i\mr{H}_1})\delta_{\be\gamma} \\
    \frac{\partial dr^{i\mr{L}_1}_\be}{\partial f(\br^{\mr{H}_1})} =&
    \left(\mr{cos}(\theta')e^{iZ}_\be - \mr{sin}(\theta')e^{iY}_\be\right) \\
    \frac{\partial dr^{i\mr{L}_l}_\be}{\partial \mr{cos}(\theta')}
    \frac{\partial \mr{cos}(\theta')}{\partial \theta'} =&
    e^{iZ}_\be\left(\mr{sin}(\theta')\right)f(\br^{i\mr{H}_1}) \\
    \frac{\partial dr^{i\mr{L}_l}_\be}{\partial \mr{sin}(\theta')}
    \frac{\partial \mr{sin}(\theta')}{\partial \theta'} =& 
    e^{iY}_\be\left(-\mr{cos}(\theta')\right)f(\br^{i\mr{H}_1})
\end{align}
and the two remaining latter terms are
\begin{align}
     \frac{\partial f(\br^{\mr{H}_1})}{\partial r^{i\mr{H}_1}_\al} =& 
     -b\frac{(\br^{\mr{O}}-\br^{\mr{H}_1})_\al}
     {\mid\br^{\mr{O}}-\br^{\mr{H}_1}\mid}
     -2cf(\br^{\mr{H}_1})\frac{(\br^{\mr{O}}-\br^{\mr{H}_1})_\al}
     {\mid\br^{\mr{O}}-\br^{\mr{H}_1}\mid} \\
     \frac{\partial \theta'}{\partial r^{i\mr{H}_1}_\al} =& 
     \frac{1}{\sqrt{1 - x^2}}\left(-\frac{(\br^{\mr{O}}-\br^{\mr{H}_2})_\al}
     {\mid\br^{\mr{O}}-\br^{\mr{H}_1}\mid\mid\br^{\mr{O}}-\br^{\mr{H}_2}\mid}
     + x \frac{(\br^{\mr{O}}-\br^{\mr{H}_1})_\al}{\mid\br^{\mr{O}}-\br^{\mr{H}_1}\mid^2}
     \right)
\end{align}
where
\begin{equation*}
    x = \frac{(\br^{\mr{O}}-\br^{\mr{H}_1})\cdot(\br^{\mr{O}}-\br^{\mr{H}_2})}
     {\mid\br^{\mr{O}}-\br^{\mr{H}_1}\mid\mid\br^{\mr{O}}-\br^{\mr{H}_2}\mid}
\end{equation*}
and
\begin{equation}
    \theta' = \mr{arccos}(x)
\end{equation}

Finally, the third term on the right hand side in the force expression of equation \ref{eq:f-tot} is
the partial derivatives of the external field at each site with
respect to the atomic positions. It is of the general form
\begin{equation}
    \frac{\partial E_\mr{in+pol}}{\partial V^j_{\byde\dots\eta}}
    \frac{\partial V^j_{\byde\dots\eta}}{\partial r^k_\delta}
    \frac{\partial r^k_\delta}{\partial r^{ia}_\al} \to
    O^i_{\byde\dots\eta}V^{i}_{\abyde\dots\eta}\frac{M^a}{M^i}
\end{equation}
where $O^i_{\byde\dots\eta}$ is the $(n-1)$th ranked moment tensor (static plus induced) 
and $V^{i}_{\abyde\dots\eta}$ the corresponding $n$th ranked external potential gradient. 
Considering the electrostatic plus induction interaction expression (see the main text), this
leads to
\begin{align}
 \frac{\partial E_\mr{in+pol}}{\partial V^j_{\byde\dots\eta}}
    \frac{\partial V^j_{\byde\dots\eta}}{\partial r^k_\delta}
    \frac{\partial r^k_\delta}{\partial r^{ia}_\al}
 =& 
 \bigg((\mi_\be(\{\br^{ia}\})+\Delta\mi_\be)V^i_\ab
 + \frac{1}{3}(\qi_\by(\{\br^{ia}\}) + \Delta\qi_\by)V^i_\aby  \nonumber \\
 &\ \ \ \ \ \ \ \ \ + \frac{1}{15}\oi_\byd V^i_\abyd + \frac{1}{105}\hi_\byde V^i_\abyde\bigg)
 \frac{M^a}{M^i}
\end{align}

\begin{figure}
\includegraphics[width=0.49\textwidth]{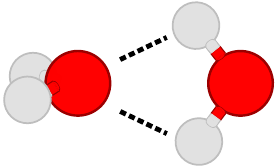}
\includegraphics[width=0.49\textwidth]{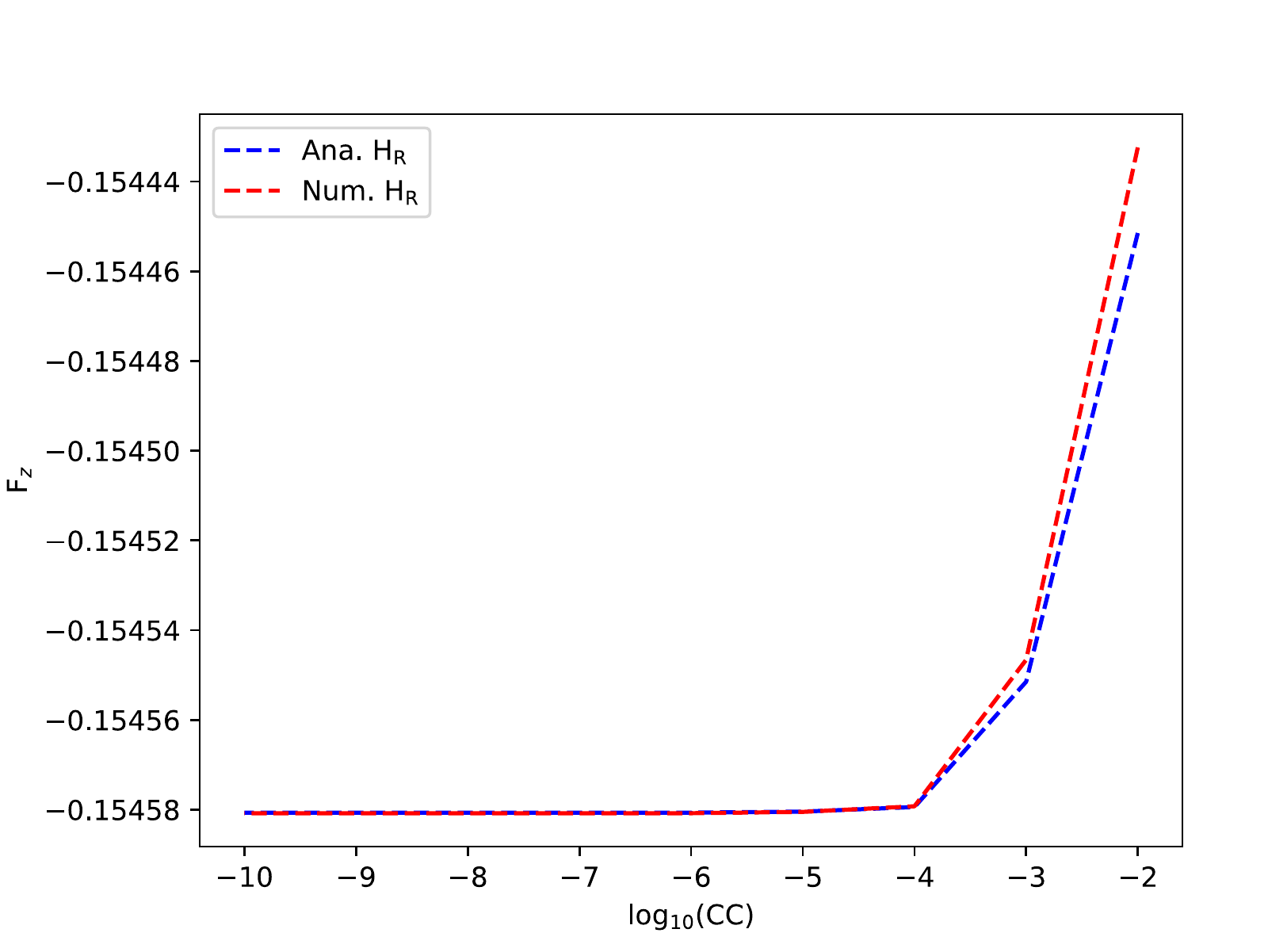}
\includegraphics[width=0.49\textwidth]{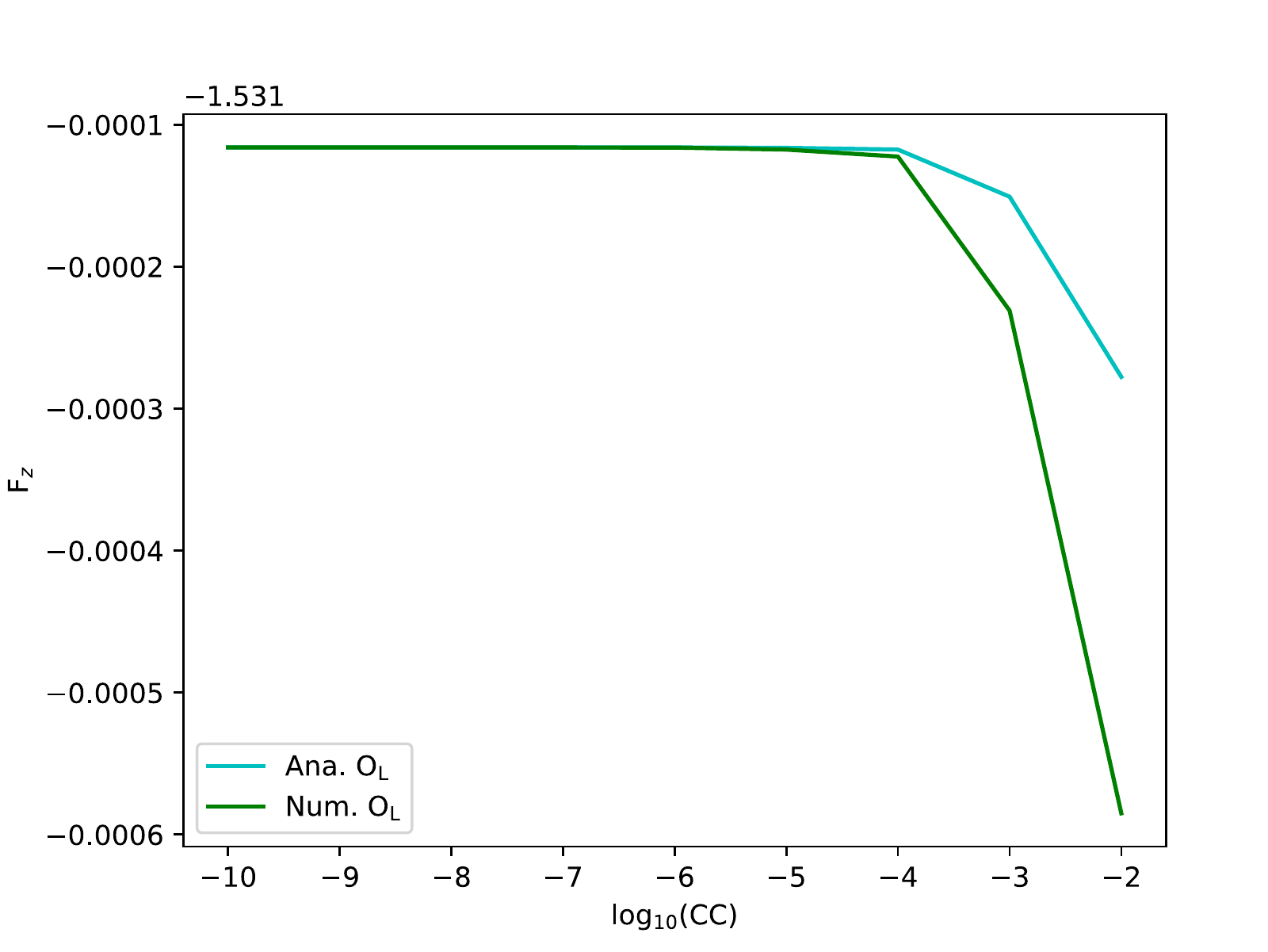}
\includegraphics[width=0.49\textwidth]{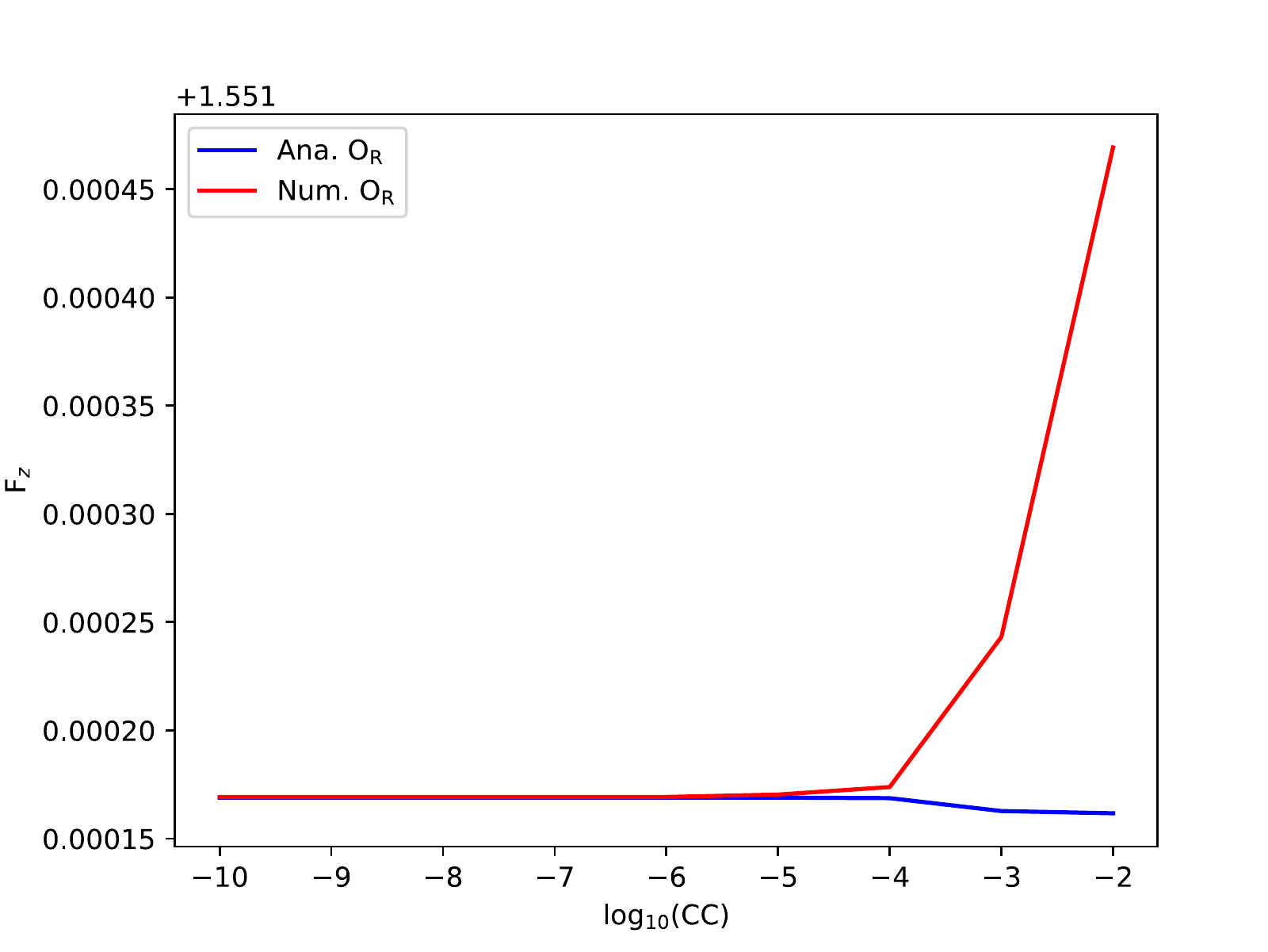}
\caption{%
C2v isomer configuration for the water dimer (top left), used for the numerical versus analytical forces at different convergence criteria ranges for one of the right hydrogens, H$_\mr{R}$ (top right), as well as the left and right oxygen, O$_\mr{L}$ (bottom left) and O$_\mr{R}$ (bottom right).
The convergence of the force components is shown versus the magnitude of the convergence criteria, CC, which is defined as $\sum_i\mid\dmi_{n+1}-\dmi_{n}\mid \leq \mr{CC}$, and similarly for the induced quadrupole moment. 
Good energy-force consistency is reached reliably at a criteria of 1e-6 D for the dipole (or D\AA\ for the quadrupole).
}
\label{fgr:force}
\end{figure}

The energy-force consistency of the formulation is checked against numerical forces, and 
at different convergence criteria of the induced moments, see figure~\ref{fgr:force}. 

\newpage

\clearpage

\section{M-site Models and RMS Deviation of M-site and QMS versus Geometric Change}

The general expression for the quadrupole moment of the three charge M-site model is
\begin{equation}
     \qi_\ab(\br^{i\mr{O}},\br^{i\mr{H}_1},\br^{i\mr{H}_2}) = \sum_{a\in i}
     ^{\mr{H}_1^{\gamma},\mr{H}_2^{\gamma},\mr{M}} 
     \frac{3}{2}\bigg\{q^{ia}\left(
     (\br^{ia}-\br^i)_\al(\br^{ia}-\br^i)_\be
     -\frac{\delta_\ab}{3}||\br^{ia}-\br^i||\right)\bigg\} 
\end{equation}
where for any finite value of $\gamma$ (i.e. placement of the M-site) 
the charges are scaled according to
\begin{equation}
    q^{\mr{H}_l^{\gamma}} = \frac{q^{\mr{H}_l}}{1-\gamma},\ \ \ \ \
    q^\mr{M} = -q^{\mr{H}_1^{\gamma}} - q^{\mr{H}_2^\gamma} \label{eq:mch}
\end{equation}
and in this way the dipole moment remains unchanged. All models make use of the optimal $\gamma$ value
of 0.4071 (see below). For the fixed charge model ($\gamma-q^\mathrm{H}_\mathrm{eq}$) the charge on the hydrogen is 0.33097$e$. Due to the relation between the placement of the M-site and the scaling of the charges
the scaled fixed charge model ($\gamma-q^\mathrm{H,*}_\mathrm{eq}$) corresponds to a shift of the position of the M-site to a $\gamma=0.3486$. Similarly the optimal $\gamma$ value is used for the M-site DMS charge model ($\gamma-\mathrm{DMS}$), and the scaled M-site DMS model ($\gamma-\mathrm{DMS}^*$) corresponds to a shift of the M-site with $\gamma=0.3838$. 

To solve for the $\gamma$ factor and hence the position of the M-site such that $\Delta$ vanishes in
\begin{equation}
    \theta = \begin{bmatrix}
             \theta_{T}-\Delta & 0 & 0 \\
             0 & -\theta_{T}-\Delta & 0 \\
             0 & 0 & 2\Delta
            \end{bmatrix} \label{eq:theta}
\end{equation}
one considers the water monomer in the equilibrium configuration with the oxygen placed at the origin.
In this configuration and frame of reference the quadrupole moment tensor only has 
components along the trace $\{\theta_{xx},\theta_{yy},\theta_{zz}\}$,
where $\theta_{zz}=2\Delta$. This component, in terms of the charges 
and positions of the hydrogens and the M-site, is given by\cite{abascal2007water}
\begin{equation}
  \theta_{zz} = q^\mr{H}(-(r_x^\mr{H})^2 + 2(r_z^\mr{H})^2 - 2(r_z^\mr{M})^2)
\end{equation}
which is trivial to solve in order for
$\Delta$ to vanish by using a generalization of the $\gamma$ expression in the global coordinate frame\cite{reimers1982intermolecular,reimers1984structure,suhm1991parameterized}
\begin{equation}
    \br_\mr{M} = (1-\gamma)\br_\mr{O} + \frac{\gamma}{2}(\br_{\mr{H}_1} + \br_{\mr{H}_2})
\end{equation}

The general expression for the $\%$RMS deviation is given by
\begin{equation}
    \% \mathrm{RMSD} = \sqrt{\frac{\sum_i^{x,y,z}(\theta_{ii} - \theta^{\mathrm{ab\ initio}}_{ii})^2}{\sum_i^{x,y,z}(\theta_{ii}^\mathrm{ab\ initio})^2}}\times 100\%
\end{equation} 
and reveals if there is a large deviation of the components in the trace of the 
quadrupole moment tensor, $\{\theta_{xx}, \theta_{yy}, \theta_{zz}\}$, relative to
the trace norm. The fixed charge M-site models $\gamma-q^\mr{H}_\mr{eq}$ and $\gamma-q^\mr{H,*}_\mr{eq}$ 
have \%RMSD of 13.0\% and 10.3\%, respectively. The variable DMS charge M-site models $\gamma-\mr{DMS}$ and $\gamma-\mr{DMS}^*$ have \%RMSD of 9.9\% and 10.2\%, respectively.
The QMS model
captures both the $\theta_T$ and the $\Delta$  
component with an average \%RMSD of 1.6$\%$ evaluated over the whole range. There is no 
systematic correlation in the deviation of the quadrupole components of the QMS models and 
the RMSD of the monomer geometry from the equilibrium geometry. The RMSD of the geometry is 
evaluated with the Kabsch algorithm.\cite{kabsch:1976}

\begin{figure}
\includegraphics[width=\textwidth]{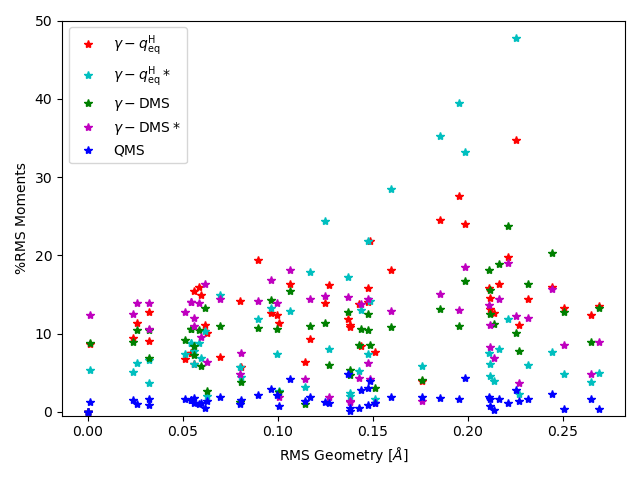}
\caption{Root-mean-square percentage difference between the three diagonal components of the quadrupole moment tensor of the M-site models and the QMS 
model versus the ab initio (ICE-CI) quadrupole moment tensor respectively. The scatter is substantial for all of the M-site models and are on average around 10$\%$. The ab initio and the average RMS percentage QMS difference
is around 1.6$\%$. The largest deviation of the QMS corresponds to the numerically lowest 
$\theta_T$. There is no correlation between the magnitude of the relative geometrical 
change relative to the ground state monomer geometry and the deviation of the quadrupole moment. 
}
\label{sfig:scatter}
\end{figure}

\clearpage

\section{Dipole moment from CCSD(T) calculations vs PS-DMS}

\begin{figure}[!ht]
\includegraphics[width=.49\textwidth]{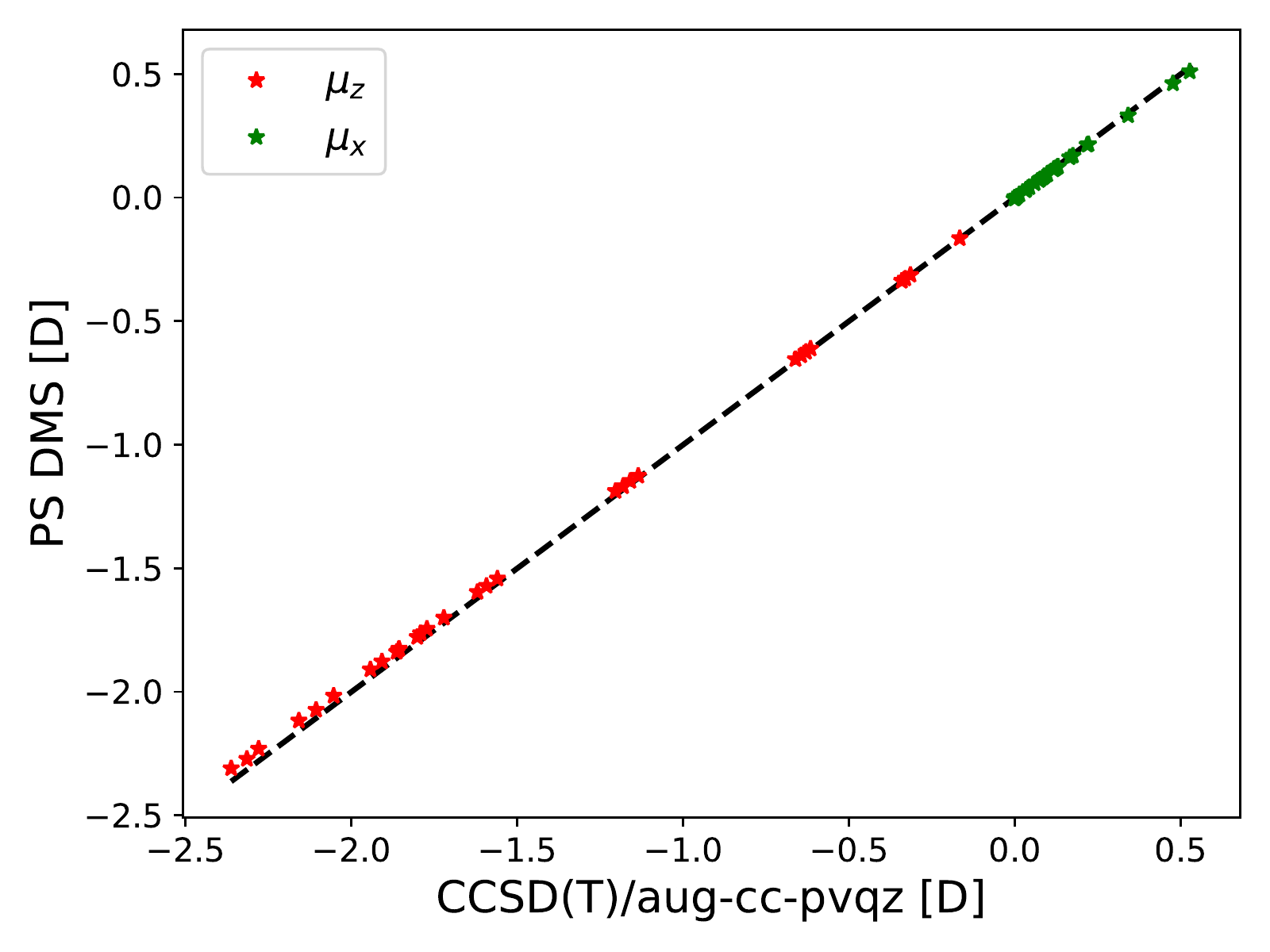} 
\includegraphics[width=.49\textwidth]{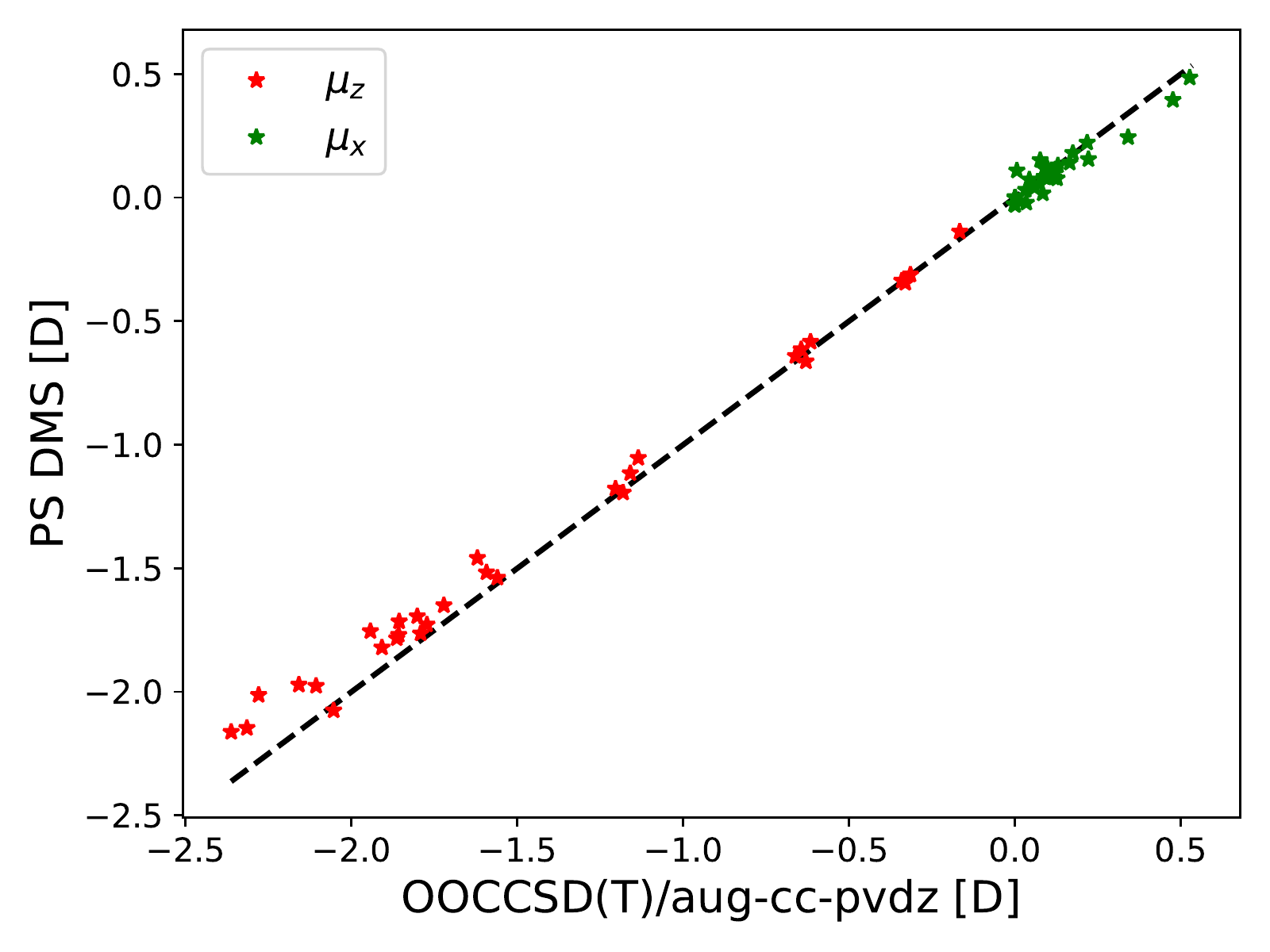}
\caption{The dipole z- and x-components, $\mu_z$ and $\mu_x$ respectively,
as predicted by CCSD(T)/aug-cc-pvqz (left, CCSD(T)) and OOCCSD(T)/aug-cc-pvdz (right, OOCCSD(T)), and 
compared to the DMS of the PS-PES. The RMSD of CCSD(T) compared to the PS DMS is 0.022 D (1.9\% difference on average), whereas 0.083 D for the OOCCSD(T) (6.9\% difference on average). For comparison the RMSD of ICE-CI 
(see main text) is 0.004 D compared to the PS DMS, or within 0.5\% on average.} 
\label{sfig:DMS-v-abinit}
\end{figure}

\clearpage

\section{Properties of Small Clusters with Fixed Quadrupole Moment}

\begin{table*}
\begin{tabular*}{\textwidth}{l@{\extracolsep{\fill}}|SSSSSSS}
\hline
\multicolumn{1}{l|}{(H$_2$O)$_n$}& 
\multicolumn{1}{c}{$E_\mr{int}$} & 
\multicolumn{1}{c}{$\Delta E_\mr{int}$} &
\multicolumn{1}{c}{$\braket{d\br_\mathrm{OO}}$} & 
\multicolumn{1}{c}{$\braket{d\br_\mathrm{OH}}$} & 
\multicolumn{1}{c}{$\braket{d\br_\mathrm{O\cdots H}}$} & 
\multicolumn{1}{c}{$\braket{d\br^a}$} & 
\multicolumn{1}{c}{$\braket{d\angle\mr{OHO}}$}
\\
\hline
  2-Cs        &  -5.05  & -0.02  & 0.022 & 0.004 & 0.023 & 0.015 & 3.784 \\ 
  3-UUD       & -15.16  & -0.21  & 0.023 & 0.017 & 0.034 & 0.040 & 2.034 \\
  4-S4        & -29.00  & -1.60  & 0.021 & 0.022 & 0.001 & 0.050 & 0.645 \\
  5-CYC       & -36.72  & -2.75  & 0.033 & 0.022 & 0.011 & 0.040 & 0.596 \\
  6-CYR       & -47.77  & -3.41  & 0.033 & 0.022 & 0.010 & 0.024 & 1.435 \\
  6-PRI       & -49.51  & -3.59  & 0.030 & 0.019 & 0.039 & 0.036 & 4.105 \\
\hline
\end{tabular*}
\caption{%
Energies and structural properties of the lowest lying water clusters including the cyclic-ring isomer of the hexamer, as predicted by the SCME/f model with a fixed quadrupole moment. 
The quadrupole moment is fixed to the moment of the ground state water monomer configuration.
$E_\mr{int}$ (kcal/mol) is the interaction energy of the clusters and $\Delta E_\mr{int}$ (kcal/mol) the difference in interaction energy compared to the CCSD(T) reference values.\cite{bates:2009,temelso:2011} 
$\braket{d\br_\mr{OO}}$, $\braket{d\br_\mr{OH}}$ and $\braket{d\br_\mr{O\cdots H}}$ are the RMSD of the oxygen-oxygen neighbour distances, intramolecular oxygen-hydrogen bond lengths of the donor-hydrogen and bonding oxygen$\cdots$hydrogen bond length distances, respectively, compared to the reference structures.
$\braket{d\br^a}$ is the overall RMSD of the relaxed SCME/f structure evaluated using the Kabsch algorithm\cite{kabsch:1976}. 
All bond related differences are in units \AA. 
$\braket{d\angle\mr{OHO}}$ is the RMSD of the angle (in degrees) between the oxygen-hydrogen-oxygen in hydrogen bonds.
}
\label{tbl:clusters}
\end{table*}

\clearpage

\section{Energy-Volume Curves for Ice Ih}

\begin{figure}[ht]
\includegraphics[width=1.0\textwidth]{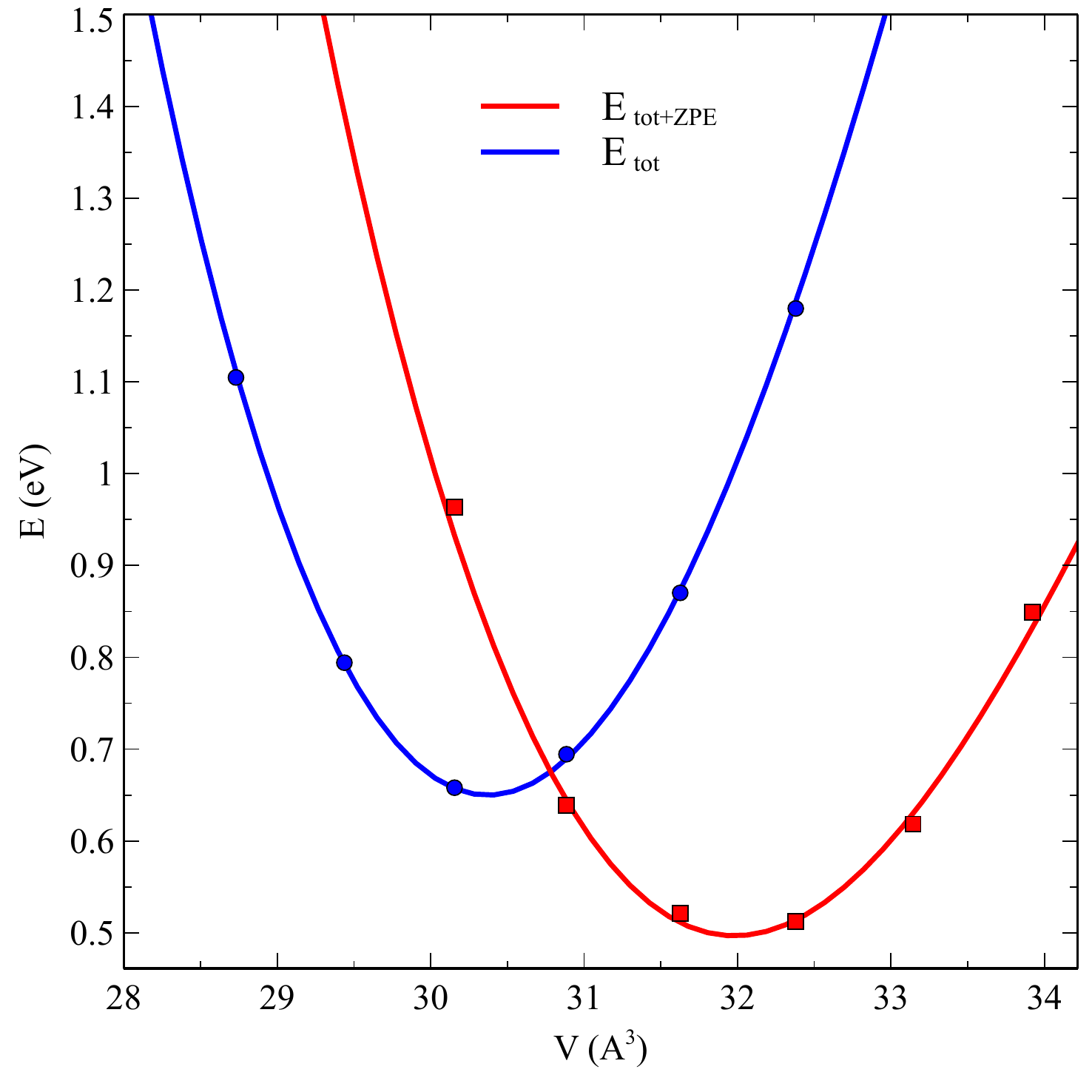}
\caption{Energy-volume curves without ($E_\mr{tot}(V)$, blue) and with zero-point-energy correction ($E_\mr{tot+ZPE}(V)$, red) obtained with the optimized (final) parameters for SCME/f.
Energies and volumes are per water molecule.}
\label{fgr:EViceIh}
\end{figure}

Figure~\ref{fgr:EViceIh} shows energy-volume curves without ($E_\mr{tot}(V)$, blue) and with zero-point-energy correction according to the quasi-harmonic approximation ($E_\mr{tot+ZPE}(V)$, red).
The indicated 6 data points are the direct results from the calculations with the optimized (final) parameters for SCME/f described in the main article around the equilibrium volume.
The lines are the results of least-square fits to the Rose-Vinet equation of state 
\begin{equation}
E(\eta) = 
E_0 \; + \;
\frac{2 \, B_0 \, V_0}{(B' - 1)^2} \cdot
\left[
2 - (5 + 3 B' \left(\eta - 1\right) - 3 \eta) \cdot 
\exp\left(-\frac{3}{2} (B'-1) \left(\eta - 1\right)\right)
\right]
\label{eq:RoseVinet}
\end{equation}
as implemented in the \textsc{PHONOPY} package \cite{togo2015},
where $\eta = \frac{V}{V_0}$ and $E_0 = -E_\mr{lat}$.
The corresponding fit parameters compiled in Table~\ref{tbl:RoseVinet}.
We have verified that adding more points does not yield significant changes for the fit parameters, in particular those that are directly compared against experimental data ($E_0$, $V_0$ and $B_0$).

\begin{table*}
    
\begin{tabular*}{\textwidth}{l@{\extracolsep{\fill}}|SSSSS}
\hline
 & 
\multicolumn{1}{c}{$E_0$ (eV)} &
\multicolumn{1}{c}{$V_0$ (\AA$^3$)} &
\multicolumn{1}{c}{$B_0$ (eV/\AA$^3$)} &
\multicolumn{1}{c}{$B_0$ (GPa)} &
\multicolumn{1}{c}{$B'$} \\
\hline
$E_\mr{tot}$ 	 				& 0.645 & 30.38 & 0.094 & 15.0 & 5.39 \\
$E_\mr{tot+ZPE}$ \hspace{3em} 	& 0.489 & 31.98 & 0.076 & 12.2 & 5.68 \\
\hline
\end{tabular*}
\caption{Fit parameters for the Rose-Vinet equation (eq~\eqref{eq:RoseVinet}) for the fits to $E_\mr{tot}(V)$ and $E_\mr{tot+ZPE}(V)$ shown in Figure~\ref{fgr:EViceIh}. All quantities are given per water molecule.
}
\label{tbl:RoseVinet}
\end{table*}

\clearpage

\bibliography{SI}